\documentclass[sigconf]{acmart}
\usepackage{subfiles}
\usepackage{makecell}
\usepackage{enumitem}
\usepackage{multirow}
\usepackage{longtable}
\usepackage{graphicx}
\usepackage{subcaption}
\usepackage{listings}
\usepackage{pgfplots}
\usepackage{pgf-pie}
\usepackage{soul}
\pgfplotsset{compat=1.17}
\usepackage{amsmath}
\usepackage{array}
\usepackage{float}
\usepackage{pdflscape}
\usepackage{rotating}
\usepackage{csquotes}
\usepackage{CJKutf8}

\newcolumntype{L}[1]{>{\raggedright\arraybackslash}p{#1}}

\newif\ifredact
\redacttrue   

\newif\ifcomment
\commenttrue   
\ifcomment
  \newcommand{\missing}[1]{\textcolor{red}{~#1}}
  \newcommand{\kel}[1]{\textcolor{olive}{~Kellie: #1}}
  \newcommand{\ken}[1]{~\sethlcolor{yellow}\hl{[Kenny: #1]}}
\else
  \newcommand{\missing}[1]{}
  \newcommand{\kel}[1]{}
  \newcommand{\ken}[1]{}
\fi

\AtBeginDocument{%
  }
\newcommand{\earnprom}{\textbf{\textit{PS1}}}
\newcommand{\bluemoon}{\textbf{\textit{PS2}}}
\newcommand{\warmsun}{\textbf{\textit{PS3}}}
\newcommand{\jaderiver}{\textbf{\textit{PS4}}}
\newcommand{\windsong}{\textbf{\textit{PS5}}}
\newcommand{\silvercove}{\textbf{\textit{PS6}}}
\newcommand{\strongarm}{\textbf{\textit{PS7}}}
\newcommand{\clearsky}{\textbf{\textit{PS8}}}
\newcommand{\calmsea}{\textbf{\textit{PS9}}}
\newcommand{\ironpetal}{\textbf{\textit{PS10}}}
\newcommand{\echoblaze}{\textbf{\textit{PS11}}}
\newcommand{\highpeak}{\textbf{\textit{PS12}}}
\newcommand{\shadowdancer}{\textbf{\textit{PS13}}}
\newcommand{\redbird}{\textbf{\textit{PS14}}}
\newcommand{\duskytide}{\textbf{\textit{PS15}}}
\newcommand{\brightstar}{\textbf{\textit{PS16}}}
\newcommand{\crystalstream}{\textbf{\textit{PS17}}}
\newcommand{\bluepea}{\textbf{\textit{PS18}}}
\newcommand{\flowymoon}{\textbf{\textit{PS19}}}
\newcommand{\butterflywave}{\textbf{\textit{PS20}}}

\newcommand{\dquote}[1]{\enquote{#1}}
\newcommand{\squote}[1]{\enquote*{#1}}

\begin{document}
\renewcommand\footnotetextcopyrightpermission[1]{} 
\pagestyle{empty} 						
\begin{CJK*}{UTF8}{gbsn}

\title{“I Said Things I Needed to Hear Myself”: Peer Support as an Emotional, Organisational, and Sociotechnical Practice in Singapore}

\author{Kellie Yu Hui Sim}
\email{kelliesyhh@gmail.com}
\orcid{0009-0005-6451-7089}
\affiliation{
  \institution{Singapore University of Technology and Design}
  \country{Singapore}
}

\author{Kenny Tsu Wei Choo}
\email{kennytwchoo@gmail.com}
\orcid{0000-0003-3845-9143}
\affiliation{
  \institution{Singapore University of Technology and Design}
  \country{Singapore}
}
\renewcommand{\shortauthors}{Sim and Choo}

\begin{abstract}
  \emph{Peer support} plays a vital role in expanding access to mental health care by providing empathetic, community-based support outside formal clinical systems. 
As digital platforms increasingly mediate such support, the design and impact of these technologies remain under-examined, particularly in Asian contexts. 
This paper presents findings from an interview study with 20 peer supporters in Singapore, who operate across diverse online, offline, and hybrid environments. 
Through a thematic analysis, we unpack how participants start, conduct, and sustain peer support, highlighting their motivations, emotional labour, and the sociocultural dimensions shaping their practices.
Building on this grounded understanding, we surface design directions for culturally responsive digital tools that scaffold rather than supplant relational care. 
Drawing insights from qualitative accounts, we offer a situated perspective on how AI might responsibly augment peer support. 
This research contributes to human-centred computing by articulating the lived realities of peer supporters and proposing design implications for trustworthy and context-sensitive AI in mental health.
\end{abstract}




\maketitle

\section{Introduction}


\emph{Peer support}, defined as the mutual provision of emotional or practical assistance between individuals with shared lived experience, has become an increasingly recognised component of mental health care~\cite{repperReviewLiteraturePeer2011, meadPeerSupportWhat2004}, providing an accessible and relational form of care for individuals who may face barriers to engaging with professional services~\cite{yeoDigitalPeerSupport2023, shahModelingMotivationalInterviewing2022, sharmaFacilitatingEmpathicConversations2021, shalabyPeerSupportMental2020, olearyDesignOpportunitiesMental2017, solomonPeerSupportPeer2004}.
Its importance is especially pronounced in non-Western or hybrid support contexts, where the boundaries between formal and informal support are fluid, and professional services may be limited, culturally mismatched, or stigmatised~\cite{kaurExploringPotentialPeer2025, sehgalExploringSocioCulturalChallenges2025, palaniappanEffectivenessPeerSupport2023, maDevelopmentMentalHealth2022, mpangoChallengesPeerSupport2020}.
Yet despite its growing prominence, peer support remains under-examined in technology design, particularly in relation to the lived realities, needs, expectations and concerns surrounding digital technologies.

The terminology surrounding peer-based interventions is diverse and inconsistently applied. 
While related to terms such as peer helping, peer counselling, and peer education, these are used variably across contexts and encompass a wide range of roles, from academic and emotional support to crisis intervention~\cite{oulanovaSuicideSurvivorPeer2014a, cowiePeerCounsellingSchools2017a, aladagEffectsPeerHelping2009a}.
These programmes differ in formality, training, and structure, often shaped by institutional norms and cultural values.
Despite definitional variation, peer support is typically distinguished from clinical services by its emphasis on mutuality, experiential knowledge, and non-hierarchical relationships~\cite{solomonPeerSupportPeer2004}.

In many Western contexts, peer support has been institutionalised through certification schemes and integrated into public mental health systems.
For instance, \textit{Certified Peer Support Specialists} in the United States and national frameworks like Australia's \textit{Peer Work Hub}\footnote{\url{https://www.nswmentalhealthcommission.com.au/content/peer-work-hub}} offer formalised training and employment pathways~\cite{shalabyPeerSupportMental2020, leePeerSupportMental2019, frankeImplementingMentalHealth2010}.
These efforts reflect a broader move towards recovery-oriented and co-produced models of care. 
However, even where formally recognised, peer support often remains peripheral to dominant biomedical paradigms, subject to institutional constraints and professional hierarchies~\cite{repperReviewLiteraturePeer2011}.

Conversely, in the Global South and many parts of Asia, peer support initiatives are more likely to be informal, volunteer-led, or embedded in educational and community settings~\cite{kaurExploringPotentialPeer2025, sehgalExploringSocioCulturalChallenges2025, palaniappanEffectivenessPeerSupport2023, maDevelopmentMentalHealth2022, mpangoChallengesPeerSupport2020}.
In Singapore, for instance, the evolving landscape reflects similar tensions.
While national programmes like the \textit{Certified Peer Support Specialist (PSS) Programme}~\footnote{\url{https://www.ssi.gov.sg/training/cet-programmes/peer-support-specialist-programme/}} gave peer support its beginnings in clinical mental health organisations, volunteer-based networks and youth-led initiatives have proliferated over recent years~\cite{SelfcareNightsPeer2024, ministryofhealthWHOLEOFSOCIETYEFFORTSSUPPORT2024, yeoDigitalPeerSupport2023, neoItsReallyNormalising2022}.
This coexistence of formal and informal approaches produces tensions around training, role clarity, and alignment with the core values of mutuality and lived experience in the peer support space.

These tensions are further complicated by the multicultural realities in Singapore, where diverse ethnic, linguistic and religious communities coexist within a densely urbanised and rapidly modernising society~\cite{frostUnsettledMajorityImmigration2021, kuahMulticulturalismSingaporeMalaysia2020, ortigaMulticulturalismItsHead2015}.
Experiences of distress, stigma, and help-seeking are mediated through historically entrenched racial policies and national values of meritocracy, self-reliance, and socio-economic productivity~\cite{frostUnsettledMajorityImmigration2021}. 
Singapore’s immigration regime has brought an influx of Chinese and Indian professionals who, while racially aligned with local majorities, are often marked as culturally different, highlighting the fragility of presumed racial commonality and complicating solidarity~\cite{ortigaMulticulturalismItsHead2015}.
These dynamics shape the social scripts through which mental illness is recognised, narrated, and addressed.
Subramaniam et al.~\cite{subramaniamStigmaPeopleMental2017} found significant differences in stigmatising attitudes across ethnic groups, suggesting that cultural and migratory legacies may intensify or mitigate stigma in ways that further complicate the mental health landscape.

Against this backdrop, there has been growing interest in the potential of digital infrastructures to support or scale peer support.
Technologies such as online forums, mobile apps, and more recently, LLMs, are being introduced to support or augment peer support, providing accessible empathetic listening, validation, and shared understanding, particularly for individuals hesitant to seek professional support~\cite{yeoDigitalPeerSupport2023, shahModelingMotivationalInterviewing2022, sharmaFacilitatingEmpathicConversations2021, olearyDesignOpportunitiesMental2017}.
More recently, LLMs have been explored for their ability to scaffold support conversations, generate emotionally attuned suggestions, or simulate distressed clients in training contexts~\cite{steenstraScaffoldingEmpathyTraining2025, yangConsistentClientSimulation2025,  wangClientCenteredAssessmentLLM2024, hsuHelpingHelperSupporting2025, sharmaHumanAICollaboration2023, sharmaFacilitatingEmpathicConversations2021}. 
However, existing tools often reflect assumptions derived from professionalised or Western models of care, potentially misaligning with the relational and culturally embedded practices of peer supporters.

Moreover, little is known about how peer supporters themselves perceive these technologies, especially in non-Western or hybrid support systems.
This lack of grounded insight is especially concerning in non-Western contexts, where support infrastructures are often fragmented, informal, or culturally specific. 
Designing responsible and context-sensitive peer support technologies thus requires a deeper understanding of the lived realities, constraints, and values that shape the work of peer supporters on the ground.
 
This study addresses these gaps by investigating the experiences and challenges of trained peer supporters engaged in volunteer-based peer support services.
Drawing on qualitative interviews with 20 participants in Singapore, we examine how they navigate emotional labour, organisational complexity, and evolving digital technologies.
Rather than assessing tools in isolation, we foreground how peer supporters interpret, adapt to, and envision the role of AI and other digital systems in their work.

Specifically, we explore the following research questions:
\begin{itemize}
    \item \textbf{RQ1}: What are the lived experiences, motivations, and challenges of trained peer supporters operating in hybrid or informal support settings?
    \item \textbf{RQ2}: How do trained peer supporters perceive the role of digital technologies, including AI, in mental health support, and what limitations and opportunities do they identify for technological interventions?
    \item \textbf{RQ3}: What support structures, training needs, and organisational conditions shape peer supporters' ability to provide support, and how might these inform the design of sociotechnical systems?
\end{itemize}

Our contributions are as follows:
\begin{enumerate}
    \item We provide empirically grounded insights into the lived experiences, motivations, and challenges of trained peer supporters operating within hybrid or informal support settings, highlighting how they navigate emotional labour, organisational complexity, and cultural expectations.
    \item We surface peer supporters’ perspectives on digital and AI-supported tools, including perceived opportunities, limitations, and tensions around authenticity, safety, and support in digitally mediated environments.
    \item We outline design considerations for the responsible integration of AI-enabled systems into volunteer-based peer support, grounded in participant reflections and situated within a non-Western context, with implications for culturally sensitive, context-aware computing design.
\end{enumerate}
\section{Related Work}
\subsection{Defining Peer Support}
Peer support refers to the mutual provision of emotional or practical assistance between individuals with shared lived experiences~\cite{meadPeerSupportWhat2004, repperReviewLiteraturePeer2011}. 
It is typically grounded in mutuality, relatability, and non-hierarchical relationships, and is often framed as a complement to formal services~\cite{solomonPeerSupportPeer2004}.
While it draws on similar communication skills as clinical care (e.g., active listening, validation), peer support usually involves less formal training and greater emphasis on lived experience~\cite{solomonPeerSupportPeer2004}.

Terminological variation (e.g., peer counselling, peer helping, peer support) further reflects divergent institutional models and cultural settings~\cite{oulanovaSuicideSurvivorPeer2014a, cowiePeerCounsellingSchools2017a, aladagEffectsPeerHelping2009a}.
In the United States and Australia, national programmes such as the \textit{Certified Peer Support Specialist} and \textit{Peer Work Hub}\footnote{\url{https://www.nswmentalhealthcommission.com.au/content/peer-work-hub}} schemes have formalised peer support roles within clinical systems~\cite{shalabyPeerSupportMental2020, leePeerSupportMental2019, frankeImplementingMentalHealth2010}. 
Elsewhere, including parts of Asia and the Global South, peer support remains informal or semi-formal, typically being integrated into school- and community-based programmes~\cite{kaurExploringPotentialPeer2025, sehgalExploringSocioCulturalChallenges2025, palaniappanEffectivenessPeerSupport2023, maDevelopmentMentalHealth2022, mpangoChallengesPeerSupport2020}.

Singapore presents a hybrid situation, where formal efforts such as the \textit{Certified Peer Support Specialist Programme} co-exist with volunteer-run and youth-led initiatives~\cite{SelfcareNightsPeer2024, ministryofhealthWHOLEOFSOCIETYEFFORTSSUPPORT2024, yeoDigitalPeerSupport2023, neoItsReallyNormalising2022}.
Yet, tensions persist around the professionalisation of peer support roles, especially in our collectivist society, where cultural norms and pervasive stigma shape how help is sought and offered~\cite{greenInfluenceCulturalNorms2024, chuaYoungAdultsMental2024, subramaniamQualitativeExplorationViews2022}.

Designing interventions for such contexts requires sensitivity not only to operational constraints but also to local expectations, values, and relational norms.
Our work builds upon this diverse landscape by foregrounding the voices of volunteer peer supporters operating in online and/or offline peer support settings, in order to surface the relational and structural tensions they navigate while taking into account culturally-specific environments.

\subsection{Online Peer Support}
The growth of digital technologies has transformed how peer support is delivered~\cite{raylandSocialNetworkPeer2023}. 
Services such as \textit{7 Cups}~\footnote{\url{https://www.7cups.com/}} and \textit{TalkLife}~\footnote{\url{https://www.talklife.com/}} facilitate text-based peer support interactions with trained volunteers.
Crisis services like \textit{Samaritans} (UK)~\footnote{\url{https://www.samaritans.org/}}, \textit{Lifeline} (Australia)~\footnote{\url{https://www.lifeline.org.au/}}, and the \textit{988 Suicide and Crisis Lifeline} (US)~\footnote{\url{https://988lifeline.org/}}, provide real-time or asynchronous support at scale, often blending professional and volunteer staffing models.
These services exemplify a shift toward more accessible, anonymous, and scalable support options.

In parallel, informal peer support communities have proliferated across platforms such as \textit{Reddit}~\cite{kimSupportersFirstUnderstanding2023, gauthierWillNotDrink2022}, \textit{Discord}~\cite{perepezkoInsteadYoureGoing2024, gohYoungPeopleCite2023}, \textit{Facebook}~\cite{yeshua-katzRoleCommunicationAffordances2021}, \textit{WhatsApp}~\cite{lambton-howardBlendingEverydayLife2021, yeshua-katzRoleCommunicationAffordances2021}, and \textit{Telegram}~\cite{yeoDigitalPeerSupport2023, rolandoTelegramSpacePeerLed2023}.
These communities are often minimally moderated and self-organised, offering flexible, low-barrier spaces where users can share mental health experiences, express their concerns, and offer mutual affirmation.
Such spaces are especially popular among youths, who value the anonymity and non-judgmental environments that these platforms afford~\cite{gohYoungPeopleCite2023}. 

Singapore reflects these global trends while exhibiting local adaptations.
Recent years have seen the emergence of a diverse ecosystem of digital mental health initiatives, including chat-based peer support, moderated online forums, and hybrid community-professional models.
Examples include \textit{SOS CareText}~\footnote{\url{https://www.sos.org.sg/pressroom/sos-launches-new-text-based-service-for-those-in-distress/}}, a primarily volunteer-operated text crisis service; \textit{let’s talk}~\footnote{\url{https://letstalk.mindline.sg}}, a moderated online forum connecting users with volunteer peer supporters and professionals; and \textit{SAFEHOUSE}~\footnote{\url{https://www.limitless.sg/safehouse}}, a private Discord community blending peer support with referral pathways, supported by both trained peer supporters and professional staff. 
While these platforms vary in their degree of formality and professional oversight, they collectively demonstrate how digital infrastructures are expanding access to mental health support outside of traditional clinical settings.

The emerging transition to online peer support offers key benefits such as accessibility, anonymity, and flexibility, making support more approachable for individuals who may feel hesitant about seeking formal or face-to-face help~\cite{kruzanInvestigatingSelfinjurySupport2021}.
These environments also foster safe, stigma-reducing environments, enabling open sharing around sensitive or distressing topics~\cite{iftikharTogetherNotTogether2023}.
Such platforms often serve as entry points for distressed individuals, providing emotional and informational support when professional services are inaccessible or stigmatised~\cite{naslundFutureMentalHealth2016}. 
Emerging evidence suggests such interventions may lead to improvements in psychological well-being and reductions in self-reported anxiety and depressive symptoms~\cite{yeoDigitalPeerSupport2023}.

Despite their benefits, online peer support platforms are not without limitations. 
Text-based communication limits access to non-verbal cues essential for conveying empathy and accurately interpreting distress~\cite {iftikharTogetherNotTogether2023, terryEmergingIssueDigital2016}. 
These cues could include tone, body language, and facial expression.
Volunteer turnover and burnout are common, and many platforms lack standardised training protocols, supervision, or feedback mechanisms~\cite{chenScaffoldingOnlinePeersupport2021}. 
There is also a risk of common techniques, such as reflective listening, paraphrasing, and open-ended questioning, backfiring if used repetitively or without contextual sensitivity, despite initial intents to support rapport-building~\cite{goldbergPsychotherapistsImproveTime2016}.

While AI-supported tools have shown promise for enhancing peer support through message suggestions, conversational scaffolding, and generating feedback~\cite{hsuHelpingHelperSupporting2025, liUnderstandingTherapeuticRelationship2024a, sharmaHumanAICollaboration2023, sharmaFacilitatingEmpathicConversations2021, tananaHowYouFeel2021}, much of this work has focused on professional therapy contexts or researcher-designed interactions.
Few studies have explored how trained volunteer peer supporters themselves make sense of such technologies, particularly in relation to their lived experiences, practices, and emotional labour.

Moreover, the design of technological tools for peer support in hybrid or non-Western systems remains underexamined in HCI and mental health design literature.
There is limited understanding of how peer supporters conceptualise their role, what tensions they encounter in emotionally demanding contexts, and how they envision digital or AI-enabled tools fitting into their practice.
This represents a significant gap, given the growing proliferation of digitally mediated peer support infrastructures and increasing interest in AI-supported care.

Our study fills this gap by examining how peer supporters interpret the role of technology and AI within their support practices, contributing much-needed empirical grounding to sociotechnical design.

\subsection{Designing for Relational and Emotional Labour}
Peer support entails significant emotional labour, the effort of managing one's own feelings and displays to support others' emotional needs~\cite{poonComputerMediatedPeerSupport2021}.
Volunteer peer supporters frequently perform such "care work" by providing empathy, active listening, and encouragement, often in high-distress situations.
Prior research shows that peers can help each other cope with these demands, with peer support being used by home care workers to regulate emotions and devise strategies for handling emotionally taxing interactions~\cite{poonComputerMediatedPeerSupport2021}.
The building of such trustful and supportive relationships is central to psychosocial support in mental health settings, and potentially highlights an underexamined aspect of relational labour.
While a large amount of past work has historically centred on support recipients rather than supporters, Kim et al.~\cite{kimSupportersFirstUnderstanding2023} revealed unique challenges faced by supporters and the critical role of relational skills in their work.

With digital technologies increasingly mediating how peer support is provided and experienced, it is crucial to examine how computer-mediated channels can offer new opportunities and challenges for managing emotional labour.
These channels have been shown to enable peer mentorship and knowledge-sharing~\cite{poonComputerMediatedPeerSupport2021}, which could potentially help to make the emotional burden more manageable.
For instance, \textit{7 Cups}~\footnote{\url{https://www.7cups.com/}} was found to attract hundreds of thousands of volunteer listeners, but those volunteers often had to develop ad-hoc coping strategies due to insufficient formal training and feedback~\cite{yaoLearningBecomeVolunteer2022}.
This gap points to design opportunities such as socio-technical supports, ranging from better peer communication tools to AI-driven mentorship, to scaffold volunteer counsellors’ skills and emotional resilience~\cite{yaoLearningBecomeVolunteer2022}.
Kim et al.~\cite{kimSupportersFirstUnderstanding2023} also found that online support environments tend to emphasise emotional support over purely informational support, underscoring the need for designs that sustain empathetic atmospheres without overburdening supporters.
Digital mediation can thus both alleviate and intensify emotional labour, by extending the reach of peer support but also stripping away in-person cues and informal debriefing rituals, demanding careful design to support peer supporters’ well-being.

At the same time, there is growing critique of AI in sensitive, relational contexts like mental health settings.
For example, general-purpose LLMs such as ChatGPT have been used to self-administer emotional support, with users drawn by their immediacy, anonymity, and accessibility~\cite{alavanzaSingaporeYouthTurn2025}. 
However, these uses fall outside clinical or peer support safeguards, raising expert concerns about their inability to intervene in crises and the risk of producing unsafe or misleading advice~\cite{louChatbotsSeeGreater2023, martinengoEvaluationChatbotdeliveredInterventions2022}.
By definition, AI chatbots “cannot feel empathy as human beings do” and may lack the nuanced emotional understanding essential for effective care interventions~\cite{boucherArtificiallyIntelligentChatbots2021}. 
LLMs such as GPT-4, while capable of producing contextually relevant emotional responses, remain fundamentally limited in aligning with human emotional dynamics, as seen in work by Huang et al.~\cite{huangApatheticEmpatheticEvaluating2024} showing noticeable misalignment in LLMs' emotional responses compared to humans. 
Although these models can simulate emotional shifts in response to situational prompts, LLMs lack emotional robustness and struggle to connect conceptually similar emotional scenarios~\cite{huangApatheticEmpatheticEvaluating2024}.
Designing for relational labour, therefore, involves a cautious integration of AI: leveraging its benefits (e.g. surfacing insights or matching peers) while ensuring that human supporters remain in the loop to provide authentic compassion and ethical oversight.

Finally, organisational structures, training practices, and cultural expectations shape how emotional labour is navigated in peer support.
Peer supporters' experiences are situated within the support organisations and communities they operate in.
For example, online and telephone helplines often train peer supporters in techniques like active listening or motivational interviewing, but studies show these trainings may be brief or under-utilised in practice~\cite{yaoLearningBecomeVolunteer2022}, and peer supporters often do not have systematic ways to receive guidelines or supervision~\cite{hsuHelpingHelperSupporting2025, chaszczewiczMultiLevelFeedbackGeneration2024}.
Without ongoing support or feedback, peer supporters might struggle with setting boundaries and coping with distressing stories, risking burnout~\cite{yaoLearningBecomeVolunteer2022}.
Prior literature highlights the need for clearer roles and more systematic training to help peer supporters manage workloads~\cite{ongImplementationPeersupportServices2023a}.
Organisational policies can also create complexity: befriending-style support models that forbid giving direct advice, while well-intentioned, might clash with the safety of peer supporters and support recipients, and even risk the supporters getting harassed~\cite {bhattacharjeeWhatsPointHaving2023}.
Bhattacharjee et al.~\cite{bhattacharjeeWhatsPointHaving2023} suggest that such mismatches stem from importing Western assumptions into a different cultural context, calling for community-grounded approaches that better fit the local norms and needs.
Peer support interventions must hence account for contextual and cultural nuances, aligned with calls by Sim and Choo~\cite{simEnvisioningAIEnhancedMental2025}.
Operational support through supervision and well-designed training should be balanced with the alignment with cultural values so that both peer supporters and support recipients can feel safe, understood and empowered.

Our work hence foregrounds these relational and structural frictions, revealing how peer supporters experience and negotiate the burdens of care within hybrid organisational and technological landscapes. 
We surface opportunities for designing contextually appropriate, emotionally intelligent support systems that centre the lived realities of peer supporters.
\section{Methods}
\subsection{Participants}
We recruited 20 participants (14 female), aged between 18 and 54 ($mean = 29.45, SD = 8.17$), each with prior experience and training in peer support and/or Psychological First Aid~\cite{shultzPsychologicalFirstAid2014a}. 
Participants reported a range of peer support experiences, from approximately 6 months to over 10 years. 
Peer support experiences varied, with 3 participants providing support online, 6 offline, and 11 in both settings.
The average experience was approximately 4.28 years across both online and offline modes. 
Those supporting others online reported 6 months to 6 years of experience; those in offline or both settings reported varied durations, including up to 10 years of experience. 

Participants were drawn from both community- and institutional-based peer support programmes or settings across Singapore.
17 participants held at least a Bachelor's degree (4 majoring in Psychology), and 3 were pursuing education in tertiary programmes (1 in a diploma programme in Psychology).
The participants also came from a variety of academic and professional fields, including engineering, design, business, education, and the arts.
While not all had formal training in psychology, many had prior experience in caregiving, community work, or leadership roles that prepared them for the emotionally attuned conversations found in peer support settings.
Table~\ref{tab:participant_summary} provides an overview of the participants' demographic backgrounds and peer support experience. 

\begin{table*}[htbp]
\centering
\small
\caption{Demographic and Peer Support Experience Details}
    \begin{tabular}{@{}p{0.7cm}p{0.9cm}p{0.9cm}p{1.5cm}p{1.3cm}p{2cm}p{3.2cm}p{4cm}@{}}
    \toprule
    \textbf{ID} & \textbf{Age} & \textbf{Gender} & \textbf{Education} & \textbf{Ethnicity} & \textbf{Employment} & \textbf{Psychology Background?} & \textbf{Experience (Type \& Duration)} \\
    \midrule
    \earnprom{} & 25--34 & Female & Bachelor's & Chinese & Full-time & Yes & Online (\textasciitilde1 year) \\
    \bluemoon{} & 45--54 & Female & Bachelor's & Chinese & Full-time & No & Both (\textasciitilde5--10 years) \\
    \warmsun{} & 35--44 & Male & Bachelor's & Chinese & Full-time & No & Offline (\textasciitilde6 years) \\
    \jaderiver{} & 18--24 & Female & Diploma & Chinese & Student & Yes & Both (\textasciitilde4 years) \\
    \windsong{} & 25--34 & Male & Diploma & Chinese & Part-time & No & Both (\textasciitilde1--2 years) \\
    \silvercove{} & 18--24 & Female & Diploma & Malay & Student & No & Both (\textasciitilde2--3 years) \\
    \strongarm{} & 18--24 & Male & Bachelor's & Indian & Student & No & Offline (\textasciitilde6 years) \\
    \clearsky{} & 25--34 & Female & Bachelor's & Chinese & Full-time & Yes & Both (\textasciitilde4 years) \\
    \calmsea{} & 18--24 & Female & Bachelor's & Chinese & Part-time & No & Online (\textasciitilde0.5 years) \\
    \ironpetal{} & 25--34 & Male & Master's & Indian & Full-time & No & Both (\textasciitilde3.5 years) \\
    \echoblaze{} & 18--24 & Female & Bachelor's & Others & Contract & Yes & Both (\textasciitilde3.5 years) \\
    \highpeak{} & 25--34 & Female & Bachelor's & Chinese & Full-time & No & Offline (\textasciitilde3 years) \\
    \shadowdancer{} & 25--34 & Female & Bachelor's & Chinese & Full-time & No & Both (>10 years) \\
    \redbird{} & 25--34 & Female & Bachelor's & Chinese & Part-time & No & Offline (\textasciitilde2--3 years) \\
    \duskytide{} & 25--34 & Female & Bachelor's & Chinese & Self-employed & No & Online (\textasciitilde6 years) \\
    \brightstar{} & 25--34 & Male & Bachelor's & Chinese & Full-time & No & Both (\textasciitilde2--3 years) \\
    \crystalstream{} & 25--34 & Female & Master's & Chinese & Full-time & No & Both (\textasciitilde3--4 years) \\
    \bluepea{} & 18--24 & Female & Bachelor's & Chinese & Student & Yes & Both (\textasciitilde6 years) \\
    \flowymoon{} & 45--54 & Female & Bachelor's & Chinese & Unemployed & No & Offline (\textasciitilde10 years) \\
    \butterflywave{} & 25--34 & Male & Bachelor's & Chinese & Full-time & No & Offline (\textasciitilde1 year) \\
    \bottomrule
    \end{tabular}
\label{tab:participant_summary}
\end{table*}

\subsection{Study Protocol} 
Each participant completed a demographic questionnaire and a semi-structured interview.
The average duration of the interviews was approximately 38 minutes and 22 seconds ($SD =$ 11 minutes and 7 seconds), with durations ranging from 25 minutes and 19 seconds to 1 hour, 14 minutes and 57 seconds.
Designed to elicit participants' reflections on their peer support experiences, the interview questions focused on common support topics, perceived adequacy of training, emotional labour, and suggestions for improving peer support structures (see Appendix~\ref{appendix:study-materials}).
Participants received a USD \$7.70 voucher upon study completion. 
All interviews were audio-recorded and transcribed using Whisper large-v3~\cite{radfordRobustSpeechRecognition2022} and manually corrected.
We then employed Braun and Clarke’s six-phase thematic analysis approach~\cite{braunUsingThematicAnalysis2006} with three researchers.
\section{Results}

\subsection{Starting Peer Support: Pathways, Motivations, and Experiences}
\subsubsection{Backgrounds and Motivations}
Participants entered peer support through diverse academic, professional, and personal pathways.
While some had formal training in psychology or healthcare, others brought insights from lived experience or caregiving. 
These varied trajectories shaped how they understood mental health, related to peers, and practised peer support.
Their motivations were equally multifaceted, ranging from personal histories of distress and recovery to a sense of civic duty, faith-based values, or aspirations for growth.
Collectively, these accounts highlight peer support as a relational and ethical practice grounded in diverse forms of knowledge.

\paragraph{\textbf{Diverse Entry Pathways}}
Participants entered peer support with varied academic, professional, and personal experiences, ranging from formal training in psychology or medicine to backgrounds in caregiving, education, and community engagement. 
This diversity shaped their understanding of mental health, communication styles, and the frameworks they employed when offering support. 
While some relied on theoretical foundations from formal education, others drew upon real-world experiences and self-reflection. 
Collectively, their trajectories demonstrate how peer support accommodates multiple forms of knowledge and benefits from a richly diverse ecosystem of care.

Several participants, such as \jaderiver{}, \clearsky{}, and \bluepea{}, were pursuing or had completed studies in psychology or related disciplines, which helped them contextualise mental health issues, recognise limitations, and engage peers with sensitivity.  
Others, including \echoblaze{} and \warmsun{}, shared how internships and continued professional development opportunities, such as diplomas, deepened their understanding of the evolving mental health landscape, including digital interventions and inter-professional collaboration.

Notably, both \warmsun{} and \crystalstream{} are medical doctors who had experiences with psychiatry-related modules during their medical studies, and were provided additional exposure to mental health care through their clinical roles and training, including structured peer support among professionals and integrated services in primary or hospital settings.
\warmsun{}, working in primary care, had received additional training through a mental health diploma and regularly engaged with psychology-trained colleagues in a dedicated mental health clinic.
\crystalstream{}, in hospital-based training, was part of an informal peer support system for junior doctors, offering one-on-one emotional support during high-stress transition periods. 
These experiences sensitised them to both systemic gaps and the emotional toll within medical environments, shaping how they engaged in broader peer support efforts.

Beyond health-related fields, participants also came from varied domains such as engineering, education, business, and the arts.
Several had experience working with children, older adults, or marginalised communities, while others held leadership, caregiving, or mentorship roles. 
These prior engagements shaped their peer support ethos, enabling them to approach conversations not merely as helpers but as listeners, bridge-builders, and companions. 
This multifaceted diversity supported a more holistic model of peer support, one that values pluralistic approaches to care.

\paragraph{\textbf{Motivations to Support Others}}
Participants' motivations for becoming peer supporters reflected a wide range of personal histories and aspirations, often shaped by past struggles, relational influences, and a desire to make a difference.
For many, personal adversity played a central role. Five participants (\bluemoon{}, \windsong{}, \strongarm{}, \duskytide{}, \brightstar{}) explicitly linked their involvement to difficult pasts or mental health experiences. 
As \bluemoon{} shared:
\begin{quote}
    \dquote{I lost my dad when I was 16 [...] at the moment in time, being 16, I don’t think you would call that a very independent age [...] I was the eldest in the family, and my mom had many years of chronic depression [...] She had no support and wasn’t very educated. She didn’t know how to regulate herself and reach out [...] [But] there were people who were very kind, like my friends, they came and 守夜 (shou ye) [stay up till midnight] with me [...] They gave 白金 (bai jin) [condolence money] [...] it’s not a lot like \$20, but I know to them it was probably their whole week’s pocket money. So I felt some warmth from that [...] at different junctures of my life, there were people who very selflessly stood up for me [...] And I felt that this is what makes the world a better place than it is.}
\end{quote}

Similarly, others found motivation in lived experiences with mental health conditions. \windsong{} reflected that their personal experiences with mental health (\dquote{anxiety and some mild depression}) helped them to \dquote{relate to the experience of sufferers well}.

Other participants expressed a desire to give back to the community (\crystalstream{}, \butterflywave{}), to explore possible education or career pathways (\clearsky{}, \echoblaze{}), or to apply their own perceived strengths to help others (\ironpetal{}, \highpeak{}). 
Motivations also stemmed from religious convictions (\shadowdancer{}, \flowymoon{}), a perceived lack of mental health advocacy (\redbird{}), and concern for friends (\jaderiver{}, \silvercove{}). 
As \redbird{} explained:
\begin{quote}
    \dquote{But now with COVID having happened and the whole [...] everyone had to really take a break and actually recognise how important mental health is. So now it’s a lot more commonplace. But at that point in time, I was a lot more involved because I believed in doing something that was still not very much advocated.}
\end{quote}

Some were prompted by a sense of inadequacy in past support attempts.
\jaderiver{} recounted:
\begin{quote}
   \dquote{It’s largely because of my experience in secondary school [...] I was taught more about being empathetic or learning how to listen to your peers [...] or how to spread happiness. I thought maybe I [could] make [my friend in distress] feel better. But [...] it was quite difficult. I didn’t have sufficient skill sets. So in Polytechnic, when I learned about this CCA, I wanted to join it [...] They said they do peer support training sessions [...] It was to learn proper skill sets to be a proper peer supporter.}
\end{quote}

\subsubsection{Lived Experiences}
Participants’ approaches to peer support were profoundly shaped by their own mental health journeys and by witnessing the struggles of those close to them. 
These experiences inspired their motivations for volunteering, deepened their empathy, and fostered a moral commitment to be present for others in distress.
For many, peer support was not merely a voluntary activity but a reciprocal practice--an attempt to extend the care they had needed or wished they had received.

\paragraph{\textbf{Firsthand Encounters with Mental Health Struggles}}
At least 13 participants, including \earnprom{}, \bluemoon{}, \warmsun{}, \jaderiver{}, \windsong{}, \silvercove{}, \strongarm{}, \clearsky{}, \echoblaze{}, \redbird{}, \duskytide{}, \brightstar{} and \crystalstream{}, described their experiences with burnout, mental health conditions such as anxiety and depression, or suicidal ideation.
These experiences, often beginning in adolescence, cultivated a sense of solidarity and informed their decision to support others.

\echoblaze{} reflected on how recovery from personal crises led to a desire to support others, stating, \dquote{I had very heavy mental issues but then I [...] came to a place of mental capacity that I can help others [...] I’m very grateful of the therapy I received that I want to give back}. 
Similarly, \strongarm{} described a pivotal experience at age 10: \dquote{I was at the verge of suiciding [...] Once I got better [...] I really understood the meaning of not being able to control yourself during such unstable times}.

\silvercove{} described a secondary school environment marked by suicidal ideation, substance misuse, and violence, sharing that \dquote{20\% of the batch tried to kill themselves [...] Our mental health was horrible [...] it was very easy to slip back into that mindset. That's why I only really talk to [those friends] when I am doing mentally okay}. 
While such conditions instilled a deep sense of responsibility, they also required strong boundaries and emotional regulation to avoid retraumatisation.

Others, like \duskytide{} and \crystalstream{}, described the cumulative toll of stigma, overwork, or family-related stress. 
\crystalstream{}, a doctor, recounted extreme overwork and emotional exhaustion: \dquote{It’s very inhumane [...] sometimes more than 40+ hours [of work] [...] most of our stress comes from that}. 
\duskytide{} linked their diagnosis to a turn toward volunteering, suggesting that \dquote{one way to [...] cope better was also to give time to other people who needed to talk}.
These narratives exemplify how care for others can be a by-product of, and response to, personal lived experiences.

\paragraph{\textbf{Secondhand Exposure and Care for Others}}
Beyond their own struggles, eight participants (\bluemoon{}, \warmsun{}, \jaderiver{}, \silvercove{}, \strongarm{}, \redbird{}, \crystalstream{}, \flowymoon{}) had supported or witnessed the suffering of people close to them.
These included caring for parents with chronic depression, supporting friends in suicidal crises, and observing the effects of stigma and neglect in school or family settings.
These secondhand experiences, observed in family, friends, classmates, and colleagues, were critical in shaping their understanding of emotional distress and the limits of informal care.
They also led participants to believe that structured peer support could offer meaningful interventions where professional services were inaccessible or insufficient.

\bluemoon{} offered a layered account of caring for their mother, who had chronic depression, as well as witnessing multiple suicides in their social circle and those around them. 
\dquote{My mom had many, many years of chronic depression [...] In secondary school, I wrote many essays on depression because I cannot understand why she’s having depression [...] I’ll be bringing her to IMH and her depression medication will cost my one month’s salary}.
Their efforts to make sense of these experiences led them to read widely about cognitive behavioural therapy and emotional boundaries.
Similarly, \warmsun{}, whose relatives had experienced depression and suicide, stated, \dquote{I feel the sadness coming to me as well [...] That was my challenge because of my family history [...] I’ve learned [...] how to step backwards and manage both their emotions and our emotions}.

\silvercove{} noted that some of their friends might be struggling with mental health conditions, stating that \dquote{they can’t regulate their emotions properly, so it causes them to act out a bit more and self-sabotage}.
Similarly, \strongarm{}, \silvercove{} and \bluepea{} described how they had supported friends or classmates in active suicidal crises or journeying through their mental health conditions.
\strongarm{} recalled classmates being followed home by counsellors and teachers to prevent harm: \dquote{[Someone] said he wanted to end his life [...] my other friend, the counsellor, and the form teacher rushed out after school and made sure he was okay}.
\bluepea{}, recalling a peer’s suicide note, said:
\begin{quote}
    \dquote{There was a guy in my class who basically sent me his entire suicide note and [...] as a 17 year old I didn’t really know how to deal with that [...] I also didn’t know [...] where this guy was [...] I called the police [...] no use [...] so I like just repeatedly called my chairperson [...] he handled it [...] he also checked in on me [...] an experience I hope I do not experience again [...] but I guess now I will be more prepared.}
\end{quote}

\paragraph{\textbf{Navigating the Cost of Care}}
Across these narratives, participants did not frame their engagement in peer support as a detached or purely voluntary activity. 
Rather, it was rooted in emotional familiarity with suffering and the lived realities of navigating distress, be it their own or others'.
These experiences fostered a distinctive balance, one marked by vigilance, moral commitment, and a readiness to hold space for others, even in ambiguity or discomfort.

Firsthand experiences offered them a deep understanding of what it means to feel unheard, unsupported, or overwhelmed. Meanwhile, secondhand encounters revealed the limits of informal support and the emotional weight of being the “go-to” person for someone else’s crisis. 
For participants like \silvercove{} and \bluepea{}, engaging in peer support was both a continuation of prior informal care and a conscious attempt to do better, to offer safety, structure, and presence where they once lacked it.

Crucially, these experiences did not make participants immune to harm, but rather heightened their sensitivity to the cost of care. 
Many described the need for boundaries, self-regulation, and ongoing reflection to avoid emotional exhaustion. 
In this sense, peer support was not simply a way of helping others, but also a mode of reclaiming agency and making meaning from difficult pasts.
It was through these experiences that participants came to understand what responsible, relational support settings could look like.

\subsection{Conducting Peer Support: Practices, Challenges and Skills}
\subsubsection{What Peer Support Involves}
Participants offered detailed accounts of what peer support looked and felt like in practice, revealing its emotional intensity, relational demands, and practical complexity.
Across school-based, community-based, and online platforms, peer support was not framed as a simple act of helping but as a situated and ongoing labour. 
Participants described moments of profound impact, such as being present during a crisis, witnessing growth in others, or learning to listen without judgment. 
Yet these experiences were interwoven with tensions like the difficulty of reading cues, managing boundaries, and sustaining care without depleting oneself.
For them, peer support was not just an act of kindness but a skilled and reflexive form of support that was rewarding, emotionally expensive, and deeply contextual.

\paragraph{\textbf{Practising Peer Support Across Contexts}}
Participants described providing peer support across diverse settings, including schools, universities, community organisations, and online platforms. 
These contexts shaped how peer support was initiated, structured, and sustained. 
Some operated within formal programmes with assigned shifts and roles, while others offered support more fluidly, responding to peers in need across digital and physical spaces.
\bluemoon{}, for instance, described extending care in a chat group for single mothers and among uniformed group peers, stating, \dquote{If it’s really necessary, I’ll schedule a call or meet them in person}.
Others, such as \duskytide{}, sustained long-term one-to-one relationships with anonymous users online: \dquote{The longest I had was like a year, almost two years [...] they would come by and talk to me}.

Yet the meaning of \squote{peer support} itself was not always clear-cut. \jaderiver{} reflected on divergent definitions:
\begin{quote}
    \dquote{For peer support [...] people have [...] different perceptions [...] I have this lecturer [...] [who told] me that that was not [...] the right kind of peer support. He just gave another definition [...] But I guess because schools nowadays, they use the word peer support in this way. So I feel like maybe there’s a mixed perception of it. I don’t really think there’s a correct, a fixed definition [...] Maybe it’s better to not mix the two up. But it’s kind of difficult for now because the industry is so complicated}.
\end{quote}
Such reflections underscore how the practice of peer support is shaped not only by context but also by competing institutional framings and individual interpretations, raising questions about who defines legitimate peer support and how these meanings evolve across settings.

\paragraph{\textbf{Challenges and Emotional Tensions}}
Participants shared that peer support could be emotionally draining, especially when they encountered peers in acute distress, or when their own emotional reserves were depleted. 
Participants described worrying about saying the wrong thing (\bluemoon{}, \jaderiver{}, \duskytide{}) or becoming burnt out from providing support (\earnprom{}, \windsong{}, \silvercove{}, \brightstar{}, \flowymoon{}).
\bluemoon{} articulated the cognitive and emotional effort involved in supporting others sensitively:
\begin{quote}
    \dquote{But I’m very worried of saying the wrong things. One, obviously, is their readiness to hear the advice. And then two is [...] you have to break it down [...] and repeat it [...] before they can hear it. Usually, you have to soothe their emotions first [...] make sure they are not triggered [...] before you can give advice. If you give advice [...] before you acknowledge their feelings, nothing is going to sink in. The challenge is a test of patience [...] you need to wait for the right moment [...] And every single word [...] you have to process it three, four times. It’s very, very draining. After that, you feel empty [...] you really have to understand [the person] [...] because it’s very easy to say something that sets people off when they are very vulnerable and fragile.}
\end{quote}

\brightstar{} similarly reflected, \dquote{It’s a lot more stressful [...] quite easy to burn out [...] I no longer have the emotional empathy to really write a proper message}.
Others, like \silvercove{}, acknowledged the tension between empathising deeply and maintaining personal wellbeing:
\begin{quote}
    \dquote{If you empathise too much, you kind of find that you are losing yourself a bit [...] you’re not really taking care of yourself, you’re just taking care of the other people around you.}
\end{quote}

Participants also expressed a sense of role ambiguity, especially when expected to offer support beyond their capacity. 
\silvercove{} noted, \dquote{If you empathise too much, you kind of find that you are losing yourself a bit [...] you’re not really taking care of yourself, you’re just taking care of the other people around you}.

Other challenges included not being able to identify whether all issues have been covered by the peer supporter (\bluepea{}), balancing between breaking confidentiality and participant safety (\brightstar{}, \bluepea{}) struggling with rapport-building and interpretation of emotional cues (\jaderiver{}, \clearsky{}, \echoblaze{}, \shadowdancer{}, \crystalstream{}), struggling with setting appropriate boundaries (\windsong{}, \silvercove{}, \highpeak{}, \flowymoon{}), and drafting non-robotic and empathetic responses (\earnprom{}, \silvercove{}, \brightstar{}).

\strongarm{} recounted the frustration of disengaged responses:
\begin{quote}
    \dquote{Sometimes they don’t entertain our questions [...] This is in primary school when I tried to talk to one of the girls. She was completely quiet [...] Even if I said hi, she wouldn’t say hi back [...] I just left it because I didn’t know what to do.}
\end{quote}

\silvercove{} offered a detailed reflection on the toll of sustained empathy, especially in digital contexts:
\begin{quote}
    \dquote{Empathy and sympathy [...] it’s also not getting tired when you do that for too much too long [...] When someone is going through something [...] it’s a bit hard to recognise that [...] Virtually, it’s almost impossible. Physically, sometimes you can tell [...] based on my facial expression and tone, they can feel the emotions I feel for them. But online, it’s hard. Not many want to video call while crying [...] It’s very hard to show emotions on text [...] I would say that I burn out quite easily [...] I wish I could help people more without tiring out as much.}
\end{quote}

\paragraph{\textbf{Navigating Difficulty and Drawing Boundaries}}
To manage these challenges, participants employed personal strategies such as drawing on training materials, seeking supervision, and setting clear boundaries. 
\duskytide{} described learning from observing their own counsellors: \dquote{I've been reading, going through different types of counsellors, also observing how they respond to me}. 
Others developed scripts for turning down requests when needed.
For example, \silvercove{} shared, \dquote{I would tell them, \squote{I’m sorry, I’m not able to be there for you [...] because I’m going through something on my own.}}.

Some used redirection or non-confrontational strategies to manage difficult conversations. Four participants (\ironpetal{}, \highpeak{}, \crystalstream{}, \butterflywave{}) shared how they supported peers without escalating tension or invalidating their perspectives.
\ironpetal{} explained:
\begin{quote}
    \dquote{I don’t know if I can say anything about it. I cannot agree, but I don’t want to disagree. so that it might trigger the participant. If I say, no, what you’re saying is just nonsense and rubbish, don’t think about it. So I wouldn’t jump to that conclusion, because probably they have been bombarded with that messaging for a very long time. And I don’t think they would change their opinion if I just tell them in a first meetup without that trust.}
\end{quote}

Participants also recognised the limits of their role and the need to manage expectations. 
\windsong{} reflected on situations where peers expressed ongoing distress without easy solutions:
\begin{quote}
    \dquote{They tend to go on and on, some of them. I don’t want to say go on and on, but then I just [...] you know, like in mental health support settings, I think some people come expecting solutions or [for supporters to be] constantly be there for them, [but] it’s not really like you can give them a solution for what they’re experiencing [...] just [need] to be constantly there to listen to them or provide a solution for them.}
\end{quote}

These narratives reflect the emotional maturity required to maintain care work over time. 
Yet, they also underscore how boundary-setting remained a largely individual responsibility, often without institutional reinforcement.
While participants actively managed the relational dynamics of peer support, their strategies to navigate challenges were shaped more by personal intuition and peer reflection than by formal structures or systemic safeguards.

\paragraph{\textbf{Common Conversational Topics}}
Participants supported peers across a wide range of concerns, from everyday stressors to severe psychological crises. 
In university and school-based contexts, academic pressure, time management, and identity-related stress were frequently cited, as mentioned by five participants (\jaderiver{}, \ironpetal{}, \echoblaze{}, \brightstar{}, \bluepea{}). 
\echoblaze{} explained, \dquote{Most sessions were about not getting along with roommates [...] academic stress [...] and a very rare third group was more heavily mental health related, like depression and anxiety}. 
\jaderiver{} similarly noted, \dquote{People expect friendship issues, but it’s mostly people with difficulties coping with school}.

Outside institutional contexts, participants often supported peers facing financial hardship, family violence, or chronic mental health struggles. 
Specifically, seven participants (\earnprom{}, \bluemoon{}, \clearsky{}, \echoblaze{}, \highpeak{}, \flowymoon{}, \butterflywave{}) highlighted that family, relationship, and emotional regulation struggles are common and often intertwined with broader existential or identity issues.
\shadowdancer{} shared, \dquote{The most common one is money not enough [...] a lot of them don’t really have families. They’re quite alone}.
Other topics mentioned by \ironpetal{}, \duskytide{} and \brightstar{} included suicidal ideation, non-suicidal self-injury, and other crisis-level disclosures, often mentioned by support recipients indirectly or only after building trust.
\duskytide{} recalled receiving messages about suicidal ideation: \dquote{It was really all quite heavy [...] a lot of queries about suicide, how to prevent thoughts of self-harm}.

\paragraph{\textbf{Duration and Sustainability}}
The duration of peer support varied widely. 
Some participants engaged in short-term, event-based or school-year initiatives, while others remained involved for years. 
Sustained engagement often depended on the presence of support structures and opportunities for reflection.

For instance, \echoblaze{} described structured rotation systems with weekly debriefs and shared responsibility.
In contrast, others like \calmsea{} felt that long shifts and the intensity of providing support on the anonymous text-based support line made the experience draining. 
Sustainability was thus shaped not only by personal motivation but also by institutional design and recognition.

Across contexts, participants (\windsong{}, \strongarm{}, \shadowdancer{}) shared that trust-building and disclosure often took time, and peers often hesitated to share due to shame, social fear, or privacy concerns.

\paragraph{\textbf{Impact and Personal Transformation}}
Despite challenges faced, many participants articulated how their experiences had been deeply rewarding. 
The impact of their contributions was sometimes only visible after considerable time had passed, but such moments affirmed their sense of purpose and the transformative potential of peer support.

Several participants described moments of affirmation and appreciation that helped sustain their motivation. 
\echoblaze{} shared, \dquote{My response got a lot of likes [...] the person even replied, thank you so much. I haven’t felt this heard before}. 
\ironpetal{} similarly reflected on how indirect appreciation still held meaning: \dquote{We have even met [...] after the program ended [...] The youth was actually sharing a lot of positive feedback about us to the counsellor}. 
Others described receiving heartfelt messages of gratitude (\earnprom{}, \clearsky{}, \ironpetal{}, \echoblaze{}, \flowymoon{}, \butterflywave{}) or witnessing a peer choose to stay alive following an encounter (\silvercove{}, \flowymoon{}).

For some, impact was not about words, but about trust sustained over time. 
\flowymoon{} explained:
\begin{quote}
    \dquote{They are just thankful. Because if you’re not good, they will not come calling you [...] The greatest feedback is when they call you and update you. The fact that they have not much friends, and they still choose to come back to you [...] it’s not easy for people to share bad things [...] So when they do, there must be a level of trust that you will not hurt them.}
\end{quote}

Others spoke about helping peers reframe their self-worth or life priorities.
\crystalstream{} reflected on helping others shift their mindsets:
\begin{quote}
    \dquote{Some of them get out of it [...] a lot of them who are fixated about, oh, I’m supposed to be successful in this [...] I chose medicine, I’m supposed to go through this [...] But after you speak to them a while, they realise actually it’s not that important. There are other things in life more important [...] a lot of doors and opportunities they never realised were there because they were so fixated.}
\end{quote}

In some cases, peer support spurred self-recognition and personal growth. 
Participants described gaining emotional resilience, communication skills, and a stronger sense of purpose ( \duskytide{}, \bluepea{}). 
For \silvercove{}, peer support became a source of meaning amid academic pressure:
\begin{quote}
    \dquote{Engineering is a very heavy workload [...] We all struggle with mental health as compared to what I see in other studies [...] I want to support my friends, so I try and support the people around me [...] Because these friends of mine who[m] I knew from secondary school, they were so bright and talented. I’m glad I am able to help them get better [...] I couldn't imagine what would happen if I had lost them [...] Now they’re doing well in tertiary institutions, they are actually impacting more people in [...] positive ways. So yeah, I think the reason I joined peer support is [...] if I'm able to make a difference in someone’s life in a positive way, like life or death kind of way, then it’s a very significant impact.}
\end{quote}

Some relationships endured well beyond the support setting. \bluemoon{} shared a story of mutual care that spanned years and reversed roles:
\bluemoon{} shared a long-term, life-spanning relationship that began with a young woman she met at a playroom:
\begin{quote}
    \dquote{I got to know her when my daughter was very young [...] she was retaking her O-[Levels] [...] had a very bad relationship with her family [...] She had an unwanted pregnancy, and I was the only one who knew. I brought her to find help [...] In the end, she decided to abort [...] Many years later, when I went through my divorce, she found out [...] cleared a room in her home and said, \squote{If you need to, this room is for you [...] }. Then I was like, \squote{ [...] you don't have to do that [...] }. And then she said something to me [...] \squote{You don’t know how important you are in my life}. But it wasn’t at the moment when it happened [...] it was many, many years after.}
\end{quote}

Other participants (\bluemoon{}, \windsong{}, \clearsky{}, \highpeak{}, \brightstar{}, \butterflywave{}) found it rewarding to provide a listening ear to those who need it.
\windsong{} reflected:
\begin{quote}
    \dquote{Not really any rewarding experiences, but it just feels good to be there for someone [...] to just provide a listening ear [...] I wish that, back when I had all these issues, someone like that was there to really understand what I'm going through [and] just listen to me.}
\end{quote}

Peer support also facilitated personal transformation and skill-building. 
Participants spoke of becoming more emotionally aware, reflective, and intentional in their communication.
\redbird{} described the experience as \dquote{a very introspective experience [...] you see how people can benefit just from a little bit of emotional support}.
Similarly, \echoblaze{} noted, \dquote{I said things I needed to hear myself [...] I was able to push down my own needs for the cause [...] which made me feel validated}.
For \redbird{} and \windsong{}, peer support was a way to ensure others would not face struggles alone, as they had.
Yet not all experiences were affirming. 
\calmsea{} offered a contrasting view, stating that the journey was \dquote{not rewarding at all [and even] draining}, serving as a reminder that peer support, while meaningful for many, can also be emotionally depleting.

\subsubsection{Providing Informal Support to Friends}
Even outside formal peer support settings, participants described offering emotional support to friends, family, and broader social circles on a wide range of topics. 
These encounters were not limited to defined shifts or institutional structures but arose organically from personal histories, interpersonal trust, and perceived availability. 

\paragraph{\textbf{Everyday Practices of Support}}
Participants described supporting their friends through academic stress, relationship issues, family dysfunction, emotional regulation challenges, and emerging mental health concerns. 
This theme arose spontaneously across many interviews, suggesting how embedded such forms of informal support was in participants’ daily lives.

\bluepea{} shared, \dquote{I do [...] make myself available to my friends [...] should they need any help or [...] just talk}.
This self-ascribed availability underscores how emotional labour frequently extended beyond scheduled shifts or designated roles. 
Similarly, \highpeak{} noted, \dquote{I’m more likely to do a lot of emotional work for my friends because we have trust and we have history and they will share more}, pointing to the ways closer interpersonal relationships often facilitated deeper disclosures and sustained involvement.

\paragraph{\textbf{Expanding the Definitions of \squote{Peer} and \squote{Support}}}
Contrary to Western definitions of peer support that emphasise structured interactions between individuals with shared lived experience, participants here adopted a more fluid interpretation when asked about their experiences supporting others. 
To some, apart from the peer support settings they volunteered in, \squote{peer} encompassed friends, juniors, or acquaintances, and support extended into everyday social life.
\silvercove{} spoke of supporting peers across age groups and family backgrounds: \dquote{I have talked to quite a few people, so they have a lot of different experiences [...] parental issues [...] sibling issues [...] }.
Shared experiences often served as a basis for trust: \dquote{I guess maybe that’s why he chose me to rant to [...] he knows that I’m able to put myself in his shoes}.

\paragraph{\textbf{Tensions of Informal Support}}
Supporting others in unresolvable or chronic situations, such as family conflict, created emotional strain for some participants.
\calmsea{} reflected, \dquote{It’s challenging when [...] they complain about their family but [...] can’t walk out on [them] [...] it’s draining for everyone [...] only can [...] empathise and [...] try not to give advice}. 
Their response, grounded in patience and empathy but avoiding giving prescriptive advice to friends, reveals the tension between presence and powerlessness that surrounds supporting one's friends.

\silvercove{} similarly reflected:
\begin{quote}
    \dquote{I would say a very difficult thing sometimes is having to empathise or sympathise with the people going through something. Empathy is okay [...] like you’ve been through it before [...] But sometimes [...] if you empathise too much, you kind of find that you are losing yourself a bit [...] you’re not really taking care of yourself, you’re just taking care of the other people around you [...] It tires you out if you do it too much or for too long [...] this doubt in your head [...] you kind of start losing your goals.}
\end{quote}

Despite the absence of formal recognition or supervision, many participants felt a strong sense of responsibility towards the friends who turned to them for support.
They often served as emotional first responders within their personal networks. 
While these efforts were frequently meaningful and impactful, they also required participants to navigate boundaries, frustration, and emotional fatigue on their own terms.

\subsubsection{Practising Peer Support: Skills, Frameworks, and Gaps}
\paragraph{\textbf{Principles and Core Skills}}
Participants described peer support as a practice grounded in emotional attunement, interpersonal tact, and careful judgement. 
Rather than offering solutions outright, they emphasised the importance of presence, containment, and validation.
Commonly cited practices included affirming emotions, avoiding unsolicited advice, respecting boundaries, and holding space for others to arrive at their own insights. 
These practices were informed by both training and lived experiences, and honed through repeated interaction.

\bluemoon{} explained the importance of emotional validation and timing:
\begin{quote}
    \dquote{It’s just one way to ensure that people don’t feel so isolated. Because when I felt isolated, I had kind people around me [...] You don’t actually need to do anything for them, you [...] just need to validate their feelings. There’s this emotional coaching system [...] The unhealthy Asian way is deny and say you shouldn’t feel this way. But the healthy way is: I accept, I validate. I think there are five steps, I forgot, but ultimately it’s about acceptance [...] When they are more regulated, then you tell them what are the potential steps they can do for themselves from your point of view. But [...] if you jump at them and tell them they’re doing it wrong, that makes them more defensive [...] Nothing can be done. I will not tell them at first contact [...] they probably just need someone to validate them and to hear them out.}
\end{quote}

Similarly, \highpeak{} emphasised a non-directive approach:
\dquote{Not invalidating, not being too eager to try to solution [...] I’m just holding space. You can’t force them to drink}.

Participants also noted the need for emotional boundaries and clarity between empathy and sympathy.
\silvercove{} stated, \dquote{I only really text them when I feel like I’m okay [...] because if I’m not doing good mentally, it’s very easy for me to get influenced [...] That’s why it’s like empathy versus sympathy}. 
The ability to withhold judgement while remaining emotionally present was seen as a defining feature of effective support.

\paragraph{\textbf{Tools, Frameworks, and Organisational Scaffolding}}
While some peer supporters relied on intuition and interpersonal sensitivity, others grounded their approach in structured conversational models or organisational frameworks. 
These included escalation protocols, phrasing guides, and structured openings and closings. 
Building rapport and setting expectations with peers were crucial factors noted by \clearsky{}, while ensuring referral protocols and appropriate boundaries were set in place was mentioned as a key practice by \echoblaze{}, \flowymoon{} and \butterflywave{}.
\earnprom{} described using standardised phrasings to suggest professional help: \dquote{You have to say, \squote{some people find it helpful to go to a professional, then what do you think?}}.

Organisational scaffolding also shaped the consistency and perceived legitimacy of support. 
Participants like \earnprom{} and \calmsea{} noted that standardised organisational protocols on clear boundaries, guidance on referrals, and protocols for documentation enabled them to act confidently while staying within their scope. 
The importance of such protocols was also highlighted by \echoblaze{}, \flowymoon{} and \butterflywave{} as practices that they would adopt by themselves despite no fixed organisational protocol for this.
However, the extent and availability of such frameworks varied greatly across settings.

\paragraph{\textbf{Training Experiences}}
Participants described a wide range of training formats, ranging from short onboarding modules and government-funded workshops to intensive community-led courses and informal mentorship. 
These trainings introduced core competencies such as PFA, listening techniques, boundary-setting, and escalation procedures. 
Role-play, experiential learning, and direct exposure to facilitators with lived experience were especially valued by participants such as \bluemoon{}, \calmsea{}, \brightstar{} and \butterflywave{}.
As \bluemoon{} noted, \dquote{When [the trainer] shares, I feel like he knows what he’s talking about}.

\brightstar{} recounted a training session that blended theory and practicals: \dquote{That was the most comprehensive, I feel}. 
\jaderiver{} described a tiered system: \dquote{Level 1 is the foundation [...] psychological first aid, basic listening skills [...] before learning more in the level 2 sessions}. 
Online platforms also offered in-app onboarding, as described by \windsong{}, who recalled learning how to recognise signs of extreme distress and when to escalate.
Others, such as \ironpetal{}, described therapeutic or creative modalities, including art therapy and empathy circles, that prompted deeper introspection and enhanced their capacity to hold space.

Nine participants (\bluemoon{}, \jaderiver{}, \windsong{}, \silvercove{}, \ironpetal{}, \echoblaze{}, \duskytide{}, \brightstar{}, \bluepea{}) highlighted that training sessions were beneficial when made modular, accessible and hybrid with topic-specific level training, with refresher courses made available for continuous upskilling.

\paragraph{\textbf{Reflections on Gaps and Design Implications}}
While training was generally appreciated, participants identified several areas for improvement. 
Participants expressed interest in training that included realistic simulations and reflective prompts rather than solely didactic content. 
\duskytide{} further highlighted the value of incorporating reflective prompts and future-oriented goal-setting, sharing that helping individuals articulate a sense of purpose, no matter how small, could serve as a powerful anchor during moments of emotional crisis. 
Simple intentions such as to \dquote{finish reading a book} or \dquote{find a cure for a disease} helped some peers regain perspective and care for themselves in the process.

Several participants felt that school-based peer programmes remained overly focused on positivity or superficial mental health messaging. 
\jaderiver{} and \strongarm{} called for more evidence-based approaches, including somatic regulation techniques and more robust case scenarios to better prepare peer supporters for real-world situations. 

\shadowdancer{}, who was recruited as a volunteer through their church, only did a \dquote{chit-chat session} during onboarding and did not go through any other screening processes. 
This raises potential for pre-screening protocols to better match volunteers with roles that suit their emotional availability and communication styles, while raising questions about the suitability of peer supporters' experiences.

\warmsun{} emphasised the need for continuing education, particularly for professionals supporting populations with comorbid mental health conditions. 
Others, such as \duskytide{}, suggested that peer supporters were required to have a baseline level of demonstrated empathy, with refresher courses offered to re-establish such levels if needed.

\subsection{Sustaining Peer Support: Self, Systems, and Futures}
\subsubsection{Caring for Self}
To sustain their involvement in emotionally demanding roles, participants developed a range of self-care practices and relied on varying degrees of external support.
For many, the ability to care for others was inseparable from the need to preserve their own emotional capacity. 
Yet self-care was rarely embedded structurally, and its uptake was uneven, highlighting both individual resilience and systemic gaps.

\paragraph{\textbf{Self-Care as Learned Necessity}}
Participants described learning to care for themselves, often through burnout or emotional depletion.
While some adopted structured self-care practices such as journaling, exercise, or spiritual grounding, others described more affective strategies, including boundary-setting, intentional rest, or emotional distancing. 
Notably, self-care was rarely institutionalised or built into the formal peer support ecosystem. 
Participants were often left to devise and manage their own strategies, highlighting a structural gap in how support roles are scaffolded and sustained.

Participants adopted a range of practices to manage their emotional wellbeing, drawing from sleep hygiene, physical routines, introspective strategies, and personal rituals. 
For many, self-care was an ongoing negotiation between emotional presence and preservation.
Several participants spoke of setting strict boundaries to preserve mental health, while others described creative or spiritual practices that helped them recover from emotionally taxing interactions. 
Notably, a few participants (\crystalstream{}, \bluepea{}, \butterflywave{}) admitted that they had not previously prioritised self-care, but shared that being asked about it during the interview prompted them to reflect on its importance and consider ways to improve their own practices.

Participants engaged in a range of self-care activities to maintain emotional equilibrium. 
For four participants, physical routines served as grounding mechanisms. 
\duskytide{} shared, \dquote{Adequate sleep is quite crucial [...] if the emotional side is too murky to wade through, at least I have this physical checklist I can tick off}.
\bluemoon{} reinforced this sentiment:
\dquote{[These are] basic essentials that [individuals] tend to neglect [...] but [they] mess up [...] functioning cycles and hormonal levels}.
\earnprom{} described a more restorative approach: \dquote{Mandatory me time [...] go for [...] yoga class or aerial class [...] slowing down and re-aligning}.
\silvercove{} described adopting habits of playing volleyball.
These forms of self-care were not purely private but served as buffers against cumulative emotional exposure.

Five participants (\earnprom{}, \windsong{}, \calmsea{}, \echoblaze{}, \brightstar{}) foregrounded the importance of deliberate emotional withdrawal when burnt out or not in an appropriate state to provide support. 
\brightstar{} noted, \dquote{The best self-care was taking a short break [...] do anything but peer support}, while \jaderiver{} explained, \dquote{If I feel tired or lethargic [...] I try not to engage in any sort of serious conversation}. 
Similarly, \redbird{} reflected, \dquote{It’s about social rest [...] spending time with myself [...] [else] I can’t save anyone as well}. 
These statements reflect an awareness of emotional limits and the importance of preserving internal capacity before engaging with others' distress.

\paragraph{\textbf{Social and Recreational Coping}}
Seven participants (\earnprom{}, \bluemoon{}, \silvercove{}, \strongarm{}, \calmsea{}, \echoblaze{}, \duskytide{}) turned to trusted friends or loved ones for decompression.
Others used religion (\bluemoon{}, \flowymoon{}), games (\silvercove{}), \echoblaze{}), or media (\ironpetal{}) as affective coping strategies. 
These activities not only offered respite but created spaces where participants could disengage from emotionally charged interactions and re-centre themselves.

\paragraph{\textbf{Gaps in Self-Care and Reflection}}
Several participants acknowledged lapses in attending to their own wellbeing. 
Three participants (\crystalstream{}, \bluepea{}, \butterflywave{}) admitted that they did not engage in regular self-care, with the interview itself prompting overdue reflection. 
Others like \clearsky{} and \highpeak{} noted that peer support was not always emotionally intense, and therefore did not necessitate elaborate self-care routines, though they still emphasised the importance of boundary-setting.
Taken together, these accounts illustrate the unevenness of self-care literacy among peer supporters and the need for structured prompts or scaffolding to surface its relevance.

\paragraph{\textbf{Structural and Relational Support Systems}}
Beyond individual self-care, participants described support systems embedded within their workplaces and organisations.
These included access to counsellors, debriefs with fellow peer supporters, mentors and staff, and institutional policies oriented around wellbeing. 
Some, like \echoblaze{}, described structured models and using colour-coded systems to signal capacity: \dquote{We had weekly meetings with our other peer counsellors and every other two weeks with a therapist [...] we will just openly say, \squote{hey, I’m feeling like a red today. Is anybody free to take my spot?}}. 
Others, such as \highpeak{} and \warmsun{}, described workplace or organisational cultures that encouraged emotional transparency and self-care through reminders, resource links, and trust-building activities across teams.

Support from fellow peer supporters was particularly significant. 
Participants such as \ironpetal{}, \shadowdancer{}, and \redbird{} noted that knowing others were trained in PFA or simply \dquote{knew the feeling} helped normalise emotional lows and validated the need to step back when overwhelmed.
In school-based settings in particular, four participants (\jaderiver{}, \silvercove{}, \strongarm{}, \echoblaze{}) highlighted that they were often supported by trained professionals such as therapists and counsellors.
\jaderiver{} emphasised the value of proximity to trained professionals: \dquote{We have direct connections to our school counsellors so they can help us if we need any help when supporting our peers}.

While five other participants (\earnprom{}, \clearsky{}, \ironpetal{}, \duskytide{}, \bluepea{}) shared that they were often supported by a staff-in-charge of the peer support programme, it was unclear if such staff were required to be trained with PFA or other principles, or if they were simply part of operational and administrative efforts.

Nonetheless, the availability and quality of support varied considerably. 
While some participants described structured, resource-rich environments, others relied on informal peer connections, staff who were not always psychologically trained, or general tokens of appreciation.
This uneven terrain of care points to broader questions around the sustainability of peer support models, particularly when emotional resilience is treated as an individual responsibility rather than a shared or institutionalised one.

\subsubsection{Structures, Systems, and Peer Support Networks}
Participants’ experiences of peer support were shaped not only by personal motivation and individual skill, but also by the organisational environments in which they operated. 
Programmes varied significantly in structure, culture, and levels of institutional support. 
While some participants described encouraging systems of supervision, peer connection, and structured reflection, others recounted challenges such as unclear protocols, administrative disorganisation, and unsafe practices.

These accounts underscore that peer support is not sustained by goodwill alone. 
Its quality and sustainability hinge on systems that foster safety, coordination, responsiveness, and care. 
Where organisations enabled flexibility, relational connection, and shared learning, participants felt supported and empowered. 
In contrast, rigid, under-resourced, or ethically compromised environments introduced emotional strain and disillusionment, sometimes prompting participants to withdraw.

\paragraph{\textbf{Interactions with Other Peer Supporters}}
Participants described how interactions with other peer supporters bolstered emotional resilience and confidence.
Five participants (\earnprom{}, \bluemoon{}, \silvercove{}, \highpeak{}, \shadowdancer{}) highlighted volunteer bonding sessions organised by their organisations, while two (\echoblaze{}, \flowymoon{}) highlighted regular meetings for case discussions and check-ins. 
These created spaces for mutual support and buffered against burnout.

Through the colour-coded system described by \echoblaze{}, peer supporters could potentially replace each other before a session: \dquote{If a peer counsellor is in a red mood, somebody has to switch out}.
\jaderiver{} highlighted opportunities to meet peer supporters beyond their own institution, signalling a nationwide push towards peer support efforts: \dquote{Peer supporters from other organisations might have different experiences from mine [...] so it was very insightful}.

\paragraph{\textbf{Organisational Barriers and Reasons for Disengagement}}
Despite positive accounts, a number of participants described organisational dysfunction, including inconsistent communication, vague role boundaries, and poor follow-up. 
These gaps eroded trust, and at times even made it difficult to deliver peer support safely and confidently.
\brightstar{} shared, \dquote{They had plans to [do group facilitation] and then after that, there was radio silence [...] it’s not really that organised}.
\calmsea{} noted superficial supervision or unhelpful staff responses, observing that \dquote{they say we can talk to them [...] but they also can’t do much [...] they’ll just say, remember to do your self-care}.

Some accounts raised more serious safety concerns. 
\redbird{} recounted a disturbing case of sexual harassment within a peer support setting, which led to them stopping their volunteering, stating, \dquote{Putting staff who were problematic with peers who were vulnerable was not the best combination [...] That was why I ended my volunteering}.
They also flagged the need for greater accountability: \dquote{There should be more [quality control] with the staff [...] more safety is always very important}.
Together, these reflections illustrate that organisational infrastructure, when absent or misaligned, can significantly undermine the emotional safety of both peer supporters and those they serve.

Other reasons for disengagement included life transitions (\brightstar{}, \calmsea{}, \earnprom{}), programme closures due to funding (\duskytide{}), and mismatched expectations (\silvercove{}), who described their experience as \dquote{boring} and unhelpful for making friends. 
These insights highlight that structural failures, in addition to personal burnout, often drive attrition.

\subsubsection{Perceptions of AI in Mental Health and Peer Support}
Participants expressed ambivalence about integrating AI into peer support. 
While AI was occasionally seen as a useful aid in low-stakes situations, many expressed concerns about its inability to grasp emotional nuance, interpret cultural context, or sustain human connection. 
These accounts reflect ongoing tensions between the scale and efficiency promised by AI and the relational, situated, and affective nature of mental health support.

\paragraph{\textbf{Conditional Use of AI in Low-Risk Scenarios}}
Four participants (\earnprom{}, \bluemoon{}, \shadowdancer{}, \flowymoon{}) suggested that AI may serve a constructive role in low-risk scenarios, where support recipients are not in immediate distress but simply in need of a listening presence. 
In such cases, AI tools could provide emotional scaffolding, streamline basic conversational flows, or maintain engagement while human support is unavailable. 
However, participants emphasised that AI should not operate in isolation, and must be embedded within systems that allow seamless escalation to human responders where needed.
\bluemoon{} outlined a scale-based approach to illustrate when AI might be appropriate:
\begin{quote}
    \dquote{Let me define it in a crisis level [...] One [...] is a situation that happened but it's not affecting the person a lot [...] Ten is when the person is [...] about to crash [...] If the person is between one to three, and can still short-circuit the emotions to do some proper thinking [...] then AI might help. But if it’s three and above [...] they’ll see the letters but won't be reading.}
\end{quote}

\paragraph{\textbf{Collaboration, Not Replacement}}
Six participants (\earnprom{}, \silvercove{}, \echoblaze{}, \highpeak{}, \crystalstream{}, \flowymoon{}) cautioned against over-reliance on AI in emotionally complex interactions. 
Rather than treating AI as a substitute for human support, they argued for human-AI collaboration. 
As \crystalstream{} put it, \dquote{The human touch still needs to be explored [...] maybe a human face behind the AI once in a while}. 
This theme underscores the importance of human oversight, particularly in contexts where trust, empathy, and containment are essential.

\paragraph{\textbf{Reducing Emotional Labour}}
Some participants viewed AI as a practical aid to ease the emotional labour of peer support. 
\ironpetal{}, \brightstar{}, \crystalstream{} and \echoblaze{} suggested that AI-generated templates or suggested replies could help structure conversations without dictating them. 
These tools would need to be editable, context-sensitive, and aligned with the peer supporter’s voice.
In parallel, \calmsea{} and \earnprom{} described using AI to reframe negative thoughts and generate affirmations for self-support, suggesting that small-scale, low-intensity applications may offer value to both peer supporters and support recipients.

\paragraph{\textbf{Cultural and Emotional Limitations}}
Despite these perceived affordances, participants raised substantive concerns about the limitations of AI in providing emotionally attuned care. 
Two participants (\flowymoon{}, \calmsea{}) critiqued existing systems for failing to understand colloquial or culturally specific language. 
\flowymoon{} described frustrations with chatbots failing to interpret colloquial language:
\begin{quote}
    \dquote{They can’t catch [...] colloquial language [...] it’s quite frustrating when they don’t [...] Sometimes, I just press [...] I want to speak to customer service. I’m sick of playing [with] this thing that gives me irrelevant answers}.
\end{quote}

Others questioned whether AI could ever offer meaningful empathy.
As \silvercove{} stated, \dquote{They can trick the human into thinking it’s empathising [...] but it’s just sympathy}, suggesting that \dquote{AI wouldn’t have the same experiences as like humans. They’ve never been through it before, then it’s like they don’t feel it}.
\echoblaze{} echoed this sentiment, comparing chatbots to human therapists: \dquote{Sometimes I don’t even want to do a breathing exercise with my therapist [...] why would I do it with the bot?}, highlighting that rapport-building is an important part of the therapeutic alliance that an AI tool may not be able to do.
These reflections suggest that current AI systems risk alienating users or undermining trust if they attempt to simulate rather than support empathy.

\paragraph{\textbf{Proposed Features and Design Suggestions}}
Participants articulated specific ideas for how AI might be integrated into peer support platforms. 
These included suggested message starters, reflection prompts, escalation protocols, and predictive text based on common conversational flows, suggested by \ironpetal{}, \brightstar{}, \crystalstream{} and \echoblaze{}.
\brightstar{} described providing suggested replies that normalise feelings or offer gentle prompts, which could be edited by supporters. 
\ironpetal{} proposed community features such as internal chats where volunteers could seek real-time advice from each other.
\highpeak{} also suggested that AI could be used to summarise information from clients, such as summarising the conversation for peer supporters.
Across these suggestions, participants stressed the importance of flexibility and discretion: AI could assist, but it must not automate the peer support process.
Overall, it was suggested that these features could be helpful for peer supporters who were \dquote{just starting out}.

\paragraph{\textbf{Shared Challenges in Human and AI Support}}
Finally, many participants observed that the limitations they ascribed to AI, such as emotional flatness or communication gaps, were also present in text-based human peer support. 
Digital environments made it harder to detect distress, sustain presence, or convey warmth.
\jaderiver{} noted, \dquote{In face-to-face, you can nod [...] [but] online, it’s difficult to express empathy}.
\bluemoon{} reflected on distraction during WhatsApp-based sessions, and \crystalstream{} observed that breaks in asynchronous conversations often disrupted rapport. 
These insights suggest that the challenges of emotional connection are not limited to AI but extend across mediated forms of care.

\subsubsection{Mental Health Landscape}
Overall, participants painted a complex picture of the local mental health landscape, suggesting that there was rising awareness nationwide yet persistent stigma, expanded ranges of services yet fragmented ecosystems, and hopeful efforts met with inertia and resistance.
These reflections surfaced critical tensions in the system and offered insight into opportunities for strengthening the mental health landscape.

\paragraph{\textbf{Persistent Stigma and Uneven Literacy}}
While 17 participants (\earnprom{}, \bluemoon{}, \warmsun{}, \jaderiver{}, \windsong{}, \clearsky{}, \echoblaze{}, \highpeak{}, \shadowdancer{}, \redbird{}, \duskytide{}, \brightstar{}, \crystalstream{}, \bluepea{}, \flowymoon{}) acknowledged increased public discourse and help-seeking, particularly among youth, stigma remained a major deterrent. 
Eight participants (\silvercove{}, \strongarm{}, \clearsky{}, \redbird{}, \brightstar{}, \bluepea{}, \flowymoon{}, \butterflywave{}) described how individuals still feared being judged by peers, employers, or institutions, especially those with more severe conditions or histories involving local psychiatric hospitals. 
As \strongarm{} noted, disclosure could be seen as a liability at work, while \brightstar{} suggested that some individuals are still not keen to seek help for fear of leaving a mark on their record.

Mental health literacy was also reported to be uneven. 
Six participants (\silvercove{}, \strongarm{}, \echoblaze{}, \highpeak{}, \redbird{}, \duskytide{}) stressed the need for deeper psychoeducation, not just for youth but also for teachers, the general public, and supporters themselves.
As \redbird{} observed, many often default to solution-giving rather than listening, highlighting a gap in training and knowledge around emotional and empathetic nuance.
\echoblaze{}, \highpeak{} and \duskytide{} also hinted at a lack of literacy in certain topics, specifically on medication taking, what to expect from therapy, and how to better support one's friends.

\paragraph{\textbf{Fragmented Systems and Inconsistent Support}}
Although mental health services have expanded in recent years, participants described them as siloed, inaccessible, or difficult to navigate. 
\bluemoon{} and \bluepea{} observed that existing peer support networks and mental health courses lacked visibility or public uptake.
\brightstar{} noted that institutional- and community-based peer support networks have been increasing nationwide, but often operate in parallel rather than in coordination, despite overlapping goals.
\strongarm{} and \butterflywave{} called for a one-stop portal that could help support recipients navigate the overwhelming range of support options.
Practical barriers, including a lack of mental health insurance (\earnprom{}), expensive therapy (\echoblaze{}, \redbird{}), and long waiting times for subsidised therapy (\bluepea{}), further restricted timely access.

The role of peer support was also critically examined. 
While valued as a relational bridge, \calmsea{} described it as a \dquote{band-aid}, insufficient without structured follow-up or referrals.
\redbird{}, \ironpetal{} and \flowymoon{} highlighted the need for clearer boundaries, professional oversight, and consistent regulation in the expanding peer support space.

\paragraph{\textbf{Supporting the Peer Supporters}}
Eight participants (\warmsun{}, \silvercove{}, \clearsky{}, \ironpetal{}, \echoblaze{}, \duskytide{}, \crystalstream{}, \flowymoon{}) raised concerns about the lack of systemic support for those in helping professions and those providing support.
Burnout, emotional labour, and isolation were recurring themes. 
As \echoblaze{} cautioned, when peer supporters were overwhelmed, the impact could cascade: \dquote{A peer supporter having a bad day ruins two people’s days}. 
These concerns underscored the need for robust emotional scaffolding, be it in terms of spaces for supervision, reflective practice, and supporting the peer supporter.

Motivational gaps were also raised. 
\shadowdancer{} and \redbird{} also highlighted a lack of incentives or motivation for people to volunteer as peer supporters.
\redbird{} felt that \dquote{a lot of the services that the facilitators are providing should [involve] [...] a stipend [due to the heavy] [...] emotional load [they are expected to carry] for free}, and suggested that it could be a reason for low numbers of peer supporters.

\paragraph{\textbf{Recommendations and Design Opportunities}}
Participants offered diverse suggestions to improve the local mental health landscape. 
These included subsidised therapy (\echoblaze{}), public testimonials to reduce stigma (\butterflywave{}), recognition schemes for volunteers (\shadowdancer{}, \redbird{}), and targeted support groups based on identity or life stage (\windsong{}, \clearsky{}).
Others proposed leveraging technology to aid focus, manage crises, or offer adaptive response suggestions, without replacing the human element.

Several participants also called for more personalised and culturally grounded approaches. 
\clearsky{} and \duskytide{} advocated for interventions attuned to local sociocultural contexts, while \shadowdancer{} argued for dismantling structural barriers like the stigma associated with psychiatric hospitals.

Collectively, these reflections emphasise that improving mental health ecosystems requires more than awareness campaigns.
It demands sustained investment in human infrastructure, thoughtful integration of technological tools, and systems that recognise the emotional realities of both peer supporters and support recipients.

\subsection{Other Volunteering Experiences}
Although this study focused on peer support in mental health contexts, many participants described broader volunteering engagements that reflected a shared ethic of supporting others.
These activities included supporting marginalised communities, mentoring youth, facilitating educational sessions, raising mental health awareness, and volunteering in non-peer support roles.
While not framed as peer support, participants reflected that such experiences were driven by similar values, such as empathy, responsibility, and community solidarity.
These experiences, shaped by formative personal experiences or long-standing interests in social good, often complemented and reinforced participants’ capacity to support others across different domains.
\section{Discussion}
Peer support represents an increasingly vital yet structurally under-supported component of mental health care systems. 
As digital interventions and AI-driven tools are introduced into these ecosystems, it is essential that such systems reflect the lived realities of those providing frontline support. 
This study contributes to human-centred computing in healthcare by detailing how peer supporters navigate emotionally complex contexts and how technology, particularly AI, might be integrated to scaffold such work. 
Through a qualitatively rich exploration, we offer implications that speak to the lived realities of peer support, while identifying meaningful opportunities for computing to augment, rather than supplant, relational care.

While earlier studies have highlighted the potential of chatbots and digital mental health tools~\cite{fitzpatrickDeliveringCognitiveBehavior2017, inksterEmpathyDrivenConversationalArtificial2018}, this study foregrounds the peer supporter’s perspective, articulating how technology must be designed not only for efficacy but for ethical fit and emotional resonance.
Drawing from participants' lived experiences, practices, and critical reflections, we synthesise key design considerations across three domains: supporting peer supporters, improving access and system integration, and responsibly deploying AI in emotionally sensitive contexts.

\subsection{Designing \textit{with} and \textit{for} Peer Supporters}
Peer supporters require tools that acknowledge the emotional labour of care, foster sustainability, and preserve relational authenticity. 
These tools should not mirror clinical training alone but respond to the realities of volunteers operating outside formal therapeutic settings, where time, training, and supervisory structures may be limited.
Our findings suggest several design opportunities.

\subsubsection{Training and Skill Development} 
Technology can complement training by simulating emotionally complex interactions. 
Scenario-based modules could scaffold increasing difficulty, support role-play with simulated client personas, and model skills such as active listening, boundary-setting, and emotional attunement. 
Such simulations may emulate client personas with dynamic emotional trajectories~\cite{steenstraScaffoldingEmpathyTraining2025, wangClientCenteredAssessmentLLM2024, louieRoleplaydohEnablingDomainExperts2024a}, providing a safe space for skill-building through structured practice. 
While previous work focuses on professional settings~\cite{yangConsistentClientSimulation2025}, our findings highlight their promise in volunteer-led peer environments.

AI-driven feedback mechanisms, potentially LLM-based, could support reflection and skills growth if designed with transparency and validation.
Chaszczewicz et al.~\cite{chaszczewiczMultiLevelFeedbackGeneration2024} showed that LLMs could provide emotionally attuned feedback and conversation strategies to improve empathy and alignment. 
This builds on work in AI systems designed to offer real-time guidance, message enhancement, and context-aware suggestions, assisting peer supporters in navigating dynamic and emotionally charged conversations~\cite{liuComPeerGenerativeConversational2024, pengExploringEffectsTechnological2020, morrisArtificiallyEmpathicConversational2018, youngRoleAIPeer2024}. 

Recent systems such as \textit{CARE}~\cite{hsuHelpingHelperSupporting2025} and \textit{HAILEY}~\cite{sharmaHumanAICollaboration2023} exemplified LLMs could support reflection and skills growth. 
\textit{CARE} provides counselling strategy suggestions based on motivational interviewing frameworks, aiding novice peer supporters during live chats. 
Similarly, \textit{HAILEY} offers just-in-time feedback to peer supporters and has been shown to significantly improve perceived empathy in conversations without diminishing supporter agency. 
These examples underscore the potential of AI not just to enhance message quality, but to support learning by doing, especially for less experienced or novice peer supporters.

Building on this, LLMs may also assist with structured coaching, drawing on principles from motivational interviewing, cognitive behavioural therapy, and active listening~\cite{sharmaFacilitatingEmpathicConversations2021, tananaHowYouFeel2021}, while balancing transparency and personalisation, key concerns echoed in user-centred design research on chatbot use in mental health contexts~\cite{yooAIChatbotsMental2025}.

\subsubsection{Wellbeing and Sustainability} 
Apart from only supporting the client or support recipient, design interventions could also proactively support the peer supporter, providing support in other ways.
This includes burnout prevention features such as pop-up prompts to pause or hand off conversations, emotion detection to flag overload, and shared dashboards to coordinate load across a peer support network. 
Platforms could also support emotional regulation through journaling tools, affirmations, or access to debrief sessions.
LLM-enabled journalling systems such as \textit{MindfulDiary}~\cite{kimMindfulDiaryHarnessingLarge2024} have shown promise for reflective processing and may be repurposed to support volunteers through guided introspection and check-ins.
Such features align with principles of augmenting human connection without displacing it, reinforcing human-AI partnerships as advocated by real-world deployments of AI in digital therapy platforms~\cite{thiemeDesigningHumancenteredAI2023}.

\subsubsection{Recognition and Retention} 
Participants reflected that they treasured the recognition and appreciation shown by those whom they supported, be it through words of affirmation or through emojis affirming their responses.
This suggests that formal recognition mechanisms such as digital badges, certificates or feedback loops can potentially be added to existing systems to acknowledge sustained emotional labour by peer supporters.
This could in turn reinforce volunteer retention and communicate the value of their contribution, reducing volunteer burnout and increasing the quality of interactions.
Prior work has highlighted how feedback-informed AI systems could empower supporters by surfacing the relational impact of their actions~\cite{thiemeDesigningHumancenteredAI2023}.

\subsection{Foregrounding Underexamined Peer Support Contexts}
Our study extends existing literature by centring the experiences of peer supporters embedded in informal and non-Western support systems, extending work that has explored peer support in Global North contexts~\cite{fitzpatrickDeliveringCognitiveBehavior2017, 
yaoLearningBecomeVolunteer2022, 
kimSupportersFirstUnderstanding2023}.
Our findings reaffirm core insights from previous research, such as the emotional labour involved in peer support, the importance of trust and boundary-setting, and the challenges of digital empathy. 
However, we surface distinct tensions and dynamics that have been underexplored in the Singaporean context in earlier work.
Notably, we found that participants often drew on highly diverse forms of knowledge, including cultural, spiritual, and experiential perspectives, to navigate their role.
Rather than adopting fixed models of mental health or peer support, they engaged in situated meaning-making and reflexive care practices, shaped by hybrid infrastructures and informal learning. 
This pluralism complicates dominant design assumptions that rely on narrowly defined competencies or standardised protocols.

Moreover, participants’ ambivalence toward AI revealed a nuanced understanding of both its potential and limitations.
While past studies have documented volunteers’ reliance on intuition or ad-hoc coping strategies~\cite{yaoLearningBecomeVolunteer2022}, our findings show how peer supporters would perceive ethically acceptable, emotionally appropriate, and culturally resonant technologies. 
Taken together, our findings reinforce the need for socio-technical systems that are contextually grounded, emotionally attuned, and inclusive of diverse epistemologies. 
They underscore that peer support is not a static role but a dynamic relationship-centric practice, continually negotiated across cultural, institutional, and emotional terrains.

\subsection{Supporting System Navigation and Access}
Beyond individual interactions, peer support tools should be embedded within broader health infrastructures to extend their impact.
Participants expressed the need for ecosystems that can better bridge gaps between peer support and professional resources, especially in times of crisis.
Tools should enable seamless escalation and triage, including AI-assisted referral systems with contextual prompts and appropriate warm handoffs. 
Pre-screening protocols may also help assess emotional readiness and skill-fit for supporters.

A centralised, filterable directory of services by accessibility, cost, urgency, identity, or concern type would improve help-seeking journeys.
AI can aid in matching individuals with peer supporters based on preferred communication styles, lived experience, or support history.

Design of any tool must be rooted in the local context.
Incorporating colloquial language databases, cultural norms, and region-specific stigma barriers is vital, as highlighted in past work~\cite{simEnvisioningAIEnhancedMental2025, greenInfluenceCulturalNorms2024, iftikharTherapyNLPTask2024, songTypingCureExperiences2024, salamanca-sanabriaExploringPerceptionsMHealth2023}.
In Singapore's context, past work has shown that levels of stigma vary across age groups and gender, with older age, male gender and lower socio-economic status being associated with more negative attitudes or perceptions~\cite{subramaniamStigmaPeopleMental2017, yuanAttitudesMentalIllness2016}. 
We imagine that platforms targeting older adults may require different modalities or intergenerational facilitation, whereas youth-facing tools might embed psychoeducation into digital campaigns or gamified formats.

\subsection{Reimagining AI for Peer Support}
While participants expressed openness to integrating AI into peer support systems, they emphasised that such tools must be used with caution and clarity of purpose. 
AI was seen as potentially helpful in low-stakes contexts, particularly for providing presence, affirmation, or simple reframing. 
This mirrors findings from Young et al.~\cite{youngRoleAIPeer2024},
who found that youth preferred AI-generated responses for everyday concerns, but strongly favoured human responses in emotionally charged or crisis situations.
In our study, participants consistently stressed that AI systems must avoid simulating empathy or engaging in emotionally intensive exchanges without human oversight, as also echoed in past work~\cite{minerKeyConsiderationsIncorporating2019}.

Participants in our study also suggested AI-powered features such as editable message templates, tone modulations, and scripts for difficult situations.
These align with design opportunities identified in the \textit{CARE} and \textit{HAILEY} systems~\cite{hsuHelpingHelperSupporting2025, sharmaHumanAICollaboration2023}. 
However, participants stressed the importance of these features being contextually relevant, and that they should be editable by users, with transparency about their source and intent.

Another proposed feature included having AI assist with summarising long conversations, with user-controlled confirmation, as also explored in past work by Kampman et al.~\cite{kampmanMultiAgentDualDialogue2024a}.
Such summarisation tools could support therapist handoffs and continuation after conversation gaps, potentially reducing cognitive load during long asynchronous exchanges.
However, such functions must maintain privacy and avoid flattening emotional nuance.

Based on past experiences with AI tools, participants were acutely aware of the difference between AI's felt empathy and simulated sympathy.
LLMs should be designed to support emotional goals such as containment, validation, and pacing, while explicitly taking into account cultural adaptation and emotional resonance.

To support appropriate reliance on AI systems, interaction designs can leverage cognitive forcing functions, such as partial explanations, to prompt critical engagement without removing human agency~\cite{dejongCognitiveForcingBetter2025}.
These strategies could clarify the AI's reasoning and scaffold human judgement, which is crucial in emotionally sensitive contexts such as peer support.
Full explanations might promote blind acceptance of suggestions, giving rise to opportunities for partial explanations (e.g., the beginning and end of a reasoning path) that can introduce a level of cognitive effort and critical engagement, reducing over-reliance without entirely withholding guidance~\cite{dejongCognitiveForcingBetter2025}.
Such approaches could help peer supporters evaluate AI outputs more reflectively, balancing support with human discretion.
It is also crucial that technology does not displace human connection. 
AI must remain a relational scaffold and not a surrogate or replacement for the human touch.

\subsection{Toward Human-Centred Peer Support Systems}
Taken together, these implications call for health computing systems that centre the emotional realities, ethical tensions, and relational demands of peer support. 
As our findings and the broader literature show, AI can augment mental health support systems when carefully embedded, but must be designed with an understanding of when to intervene or when to defer to human input.
Across recent HCI research, we observe a shift from AI being a decision-maker to AI being a collaborative partner~\cite{swingerTheresNoTEAMMAIT2025a}, emphasising calibrated trust, user agency, and contextual appropriateness. In parallel, studies employing participatory and co-design methods~\cite{poulsenCodesigningConversationalGenerative2025, yooAIChatbotsMental2025} reinforce the value of involving peer supporters and communities directly in the development of AI tools, ensuring cultural alignment and emotional resonance.
Expert-guided designs can further enhance realism and alignment with established counselling practices~\cite{louieRoleplaydohEnablingDomainExperts2024a}.
By integrating AI cautiously and responsively into peer support systems, designers can amplify care while preserving the integrity and dignity of human connection.
\section{Limitations and Future Work}
This study provides a foundational, empirically grounded understanding of peer support work, highlighting implications for the design of context-aware, ethically grounded sociotechnical systems in mental health.
However, several limitations warrant consideration.

Our findings are based on a qualitatively rich but non-representative sample of peer supporters, many of whom were recruited from community and educational settings in Singapore.
This cultural specificity enables deep insight into localised practices and challenges often neglected in global HCI. 
However, it may also limit generalisability and transferability to other geographies or populations with differing mental health infrastructures, cultural norms, or peer support paradigms. 
Future research may explore comparative cross-cultural or transnational analyses to examine how peer support values and practices vary across regions, particularly in under-represented Global South contexts, and how these differ from dominant models in Global North settings.

Although participants reflected on real-world peer support experiences, the study did not directly observe interactions within live peer support contexts. 
As such, insights about technology use remain exploratory and reflective.
However, complementary work involving expert interviews and evaluations of LLM-supported prototypes has begun to explore these questions in greater detail.
This study provides the design rationale and empirical grounding for those ongoing efforts. 

Though this study foregrounds peer supporters' voices, it risks omitting the perspectives of those receiving support. 
Understanding how support recipients interpret peer supporter behaviours, emotional presence, or AI-generated content could yield important insights into relational alignment and trust.
Future research should investigate both sides of the peer support interaction, particularly in technology-mediated settings.

Given the rapid evolution of LLMs, the capabilities, risks and implications of AI in peer support settings may change significantly. 
Our findings reflect perceptions and design considerations based on current-generation tools.
Future research could adopt a living evaluation approach, revisiting user perceptions and system affordances as new models emerge and as AI integration becomes more mainstream in mental health ecosystems.

Finally, while we identified opportunities for technology to scaffold training and the infrastructure of peer support, the long-term implications of integrating AI into such settings remain underexplored. 
Future research could examine the impact of such systems on volunteer retention, emotional burden, boundary management, and support outcomes over time. 
Organisational and policy-level studies could also be considered to understand how institutional culture, governance, and incentive structures mediate the adoption of AI-augmented peer support.
\section{Conclusion}
Volunteer-based peer support is increasingly positioned as a critical supplement to overburdened mental health systems, yet it remains emotionally intensive, structurally uneven, and often undervalued.
AI-supported tools present new possibilities for scaffolding this labour, but must be designed with care to respect the relational and ethical complexity of peer support practice.

This formative study traced how peer supporters start, conduct, and sustain peer support across contexts.
In starting peer support, we found that participants’ diverse motivations and pathways brought together lived experience, academic knowledge, and caregiving histories. 
These shaped how they conceptualised care and enacted their roles, suggesting that future systems must accommodate plural definitions of expertise and support.

In conducting peer support, participants described emotionally charged, highly contextual interactions that demanded presence, tact, and continuous reflection. 
Despite past training experiences and lived experiences guiding their conversations, much of this was considered to be improvised and affectively demanding. 
Here, participants saw potential for AI to offer low-intensity support, such as conversation prompts or reflection tools, but rejected any role that displaced human discretion or emotional attunement.

In sustaining peer support, we observed significant gaps in institutional and systemic support for peer supporters.
Burnout, ambiguous role expectations, and inconsistent support infrastructures were common.
While participants experimented with individual self-care strategies, the absence of systemic reinforcement raised concerns about long-term sustainability. 
Some imagined AI-supported features that could buffer emotional labour or offer real-time peer feedback, but these would need to be culturally grounded, emotionally realistic, and aligned with lived experiences and real-world best practices.

We argue that AI should not replace, but instead extend the emotional and reflective capacity of peer supporters. 
Designing such systems requires sensitivity to local contexts, affective labour, and the often invisible infrastructures of informal care.
While this paper offers a formative contribution, we have begun to build on these insights through the development and testing of AI tools, with the goal of advancing systems that respect and sustain the ethics of peer support.
In doing so, we contribute to efforts to embed AI in emotionally sensitive domains through situated, human-centred design. 
The challenge is not simply what AI can do, but how it can be integrated to preserve the human touch and uphold the relational, ethical, and situated nature of peer support.

\begin{acks}
This research is supported by the Ministry of Education, Singapore, under its SUTD Kickstarter Initiative (SKI 2021\_04\_08).
Any opinions, findings and conclusions or recommendations expressed in this material are those of the author(s) and do not reflect the views of the Ministry of Education, Singapore.
We would like to thank our reviewers for their thoughtful feedback and constructive suggestions, which helped strengthen this work.
We would also like to thank Hugo Chen for his assistance with cleaning the interview transcripts and the subsequent data analysis, and Chenyu Zhao for her assistance with the data analysis of these transcripts.
Their contributions were invaluable to the timely and rigorous completion of the study.
We are also deeply grateful to our participants for sharing their time, experiences, and candid reflections with us.
\end{acks}

\bibliographystyle{ACM-Reference-Format} 
\bibliography{main}

\onecolumn
\appendix
\section{Study Materials}
\label{appendix:study-materials}
\subsection{Demographics Questionnaire}
The questionnaire in Table~\ref{tab:phase-1-demographics-questionnaire} collected basic demographic details from participants at the start of the study.

\begin{table}[htbp]
    \caption{Demographics Questionnaire Items}
    \centering
    \small
    \begin{tabular}{p{3.5cm} p{11cm}}
    \hline
    \textbf{Question} & \textbf{Response Options} \\
    \hline
    Age & 18–24, 25--34, 35--44, 45--54, 55--64, 65--74, 75--84, 85 and above \\
    Gender & Male, Female, Non-binary, Other \\
    Education (Current or Highest Attained) & Secondary School and Below, Polytechnic/Junior College/Institute of Technical Education/Millennia Institute, Bachelor's Degree, Master's Degree, Doctoral Degree, Other \\
    Ethnicity & Chinese, Malay, Indian, Other \\
    Employment Status & Full-time, Part-time, Contract/Temporary, Retired, Unemployed, Unable to work, Student, Other \\
    \hline
    \end{tabular}
    \label{tab:phase-1-demographics-questionnaire}
\end{table}

\subsection{Experience Interview}
The semi-structured interview guide in Table~\ref{tab:phase-1-interview-guide} was used to explore participants’ backgrounds, volunteering experiences, peer support reflections, and challenges faced in their roles.

\begin{table}[htbp]
    \centering
    \small
    \caption{Semi-Structured Interview Guide: Experience Interview}
    \begin{tabular}{p{3.5cm}p{11cm}}
    \hline
    \textbf{Section} & \textbf{Interview Questions} \\
    \hline
    General &
    1. Do you work in the mental health industry? If not, what field of work do you work in? \newline
    2. How do you think societal attitudes towards mental health have changed over recent years? \\
    Volunteering Experiences &
    1. How long have you been volunteering with [organisation]? \newline
    2. Do you volunteer with other organisations too? \newline
    3. What does your role as a volunteer usually entail? \newline
    4. What motivated you to start volunteering in the mental health field? \newline
    5. What kind of training did you receive before starting your volunteer work? \\
    Peer Support Experiences &
    1. What are some common topics that you encounter when providing peer support? \newline
    2. Can you describe a particularly rewarding experience you've had while volunteering? \newline
    3. Could you share more about your experiences chatting with clients in physical peer support settings? \newline
    4. Could you share more about your experiences chatting with clients in virtual peer support settings (i.e., through video calls or phone calls, chat-based platforms etc)? \newline
    5. Are there any challenges you faced while talking to clients physically? \newline
    6. Are there any challenges you faced while talking to clients virtually? \newline
    7. How do you handle challenging situations or crises while volunteering? \newline
    8. What is one thing that you enjoy about providing peer support, and why? \newline
    9. What is one thing that you’d like to change regarding providing peer support, and why? \newline
    10. Can you share any feedback from individuals you've supported? \\
    Self-Care and Support &
    1. What kind of support do you receive from [organisation] while volunteering? \newline
    2. How do you manage self-care while volunteering in a mentally demanding role? \\
    Miscellaneous &
    1. Are there any other challenges you face as a peer support volunteer? \\
    \hline
    \end{tabular}
    \label{tab:phase-1-interview-guide}
\end{table}

\end{CJK*}
\end{document}